\documentclass[9pt,a4paper,fleqn,twocolumn]{article}

\usepackage[utf8]{inputenc}    % french keyboard
\usepackage[english]{babel}
\usepackage[T1]{fontenc}
\usepackage{amsmath}
\usepackage{amsfonts}
\usepackage{amssymb}
\usepackage{graphicx}   % Including figure files
\usepackage{lmodern}           % modern fonts (allowing search in manuscript)
\usepackage[left=1.3cm,right=1.3cm,top=1.5cm,bottom=1.5cm]{geometry}
\graphicspath{{/home/user-777/These/Ecrits/Flyby_April2016/Figures/}}
\usepackage[round, sort&compress]{natbib}

\usepackage{array} 
\usepackage{rotfloat}          % rotate floats
\usepackage{multicol,rotating} % Columns and floats
\usepackage{subcaption}        % multi figures
\usepackage{dblfloatfix}       % allow placing of floats at pages bottom
\usepackage[section]{placeins} % set/force location of suppl. float page at the current section
\usepackage{xcolor}            % used for colored symbols
\usepackage{float}             % create a minipage as a float environment
\usepackage{hyperref}
\hypersetup{colorlinks=true, linkcolor=blue, citecolor=blue, urlcolor=blue}

\newcommand{\dgr}{$^{\circ}$ $ $}
\newcommand{\dg}{$^{\circ}$}

\title{ROSETTA/OSIRIS observations of the 67P nucleus during the April 2016 
       flyby: high-resolution spectrophotometry}
\date{accepted in A\&A, DOI:10.1051/0004-6361/201833807}
\begin{document}
\twocolumn[{%
\maketitle
\begin{center}
%%%%% Core Authors %%%%%
\parbox{\textwidth}{
\begin{normalsize}
C. Feller$^{1}$ [Corresponding author: clement.feller-at-obspm.fr], 
S. Fornasier$^{1}$, S. Ferrari$^{2}$, P.H. Hasselmann$^{1}$ , A. Barucci$^{1}$, 
M. Massironi$^{2}$, J.D.P Deshapriya$^{1}$
%%%%% Instr. PIs %%%%%%%
, H. Sierks$^{3}$, G. Naletto$^{4,5,6}$, P. L. Lamy$^{7}$, R. Rodrigo$^{8,9}$, D. Koschny$^{10}$, 
B.J.R. Davidsson$^{11}$
%%%%% Instr. Co-Is %%%%%
, J.-L. Bertaux$^{7}$, I. Bertini$^{4}$, D. Bodewits$^{12}$, G. Cremonese$^{13}$, V. Da Deppo$^{6}$, 
S. Debei$^{14}$, M. De Cecco$^{15}$, M. Fulle$^{16}$, P. J. Guti{\'e}rrez$^{17}$, C. G{\"u}ttler$^{3}$, 
W.-H. Ip$^{18,19}$, H. U. Keller$^{20,21}$, L. M. Lara$^{17}$, M. Lazzarin$^{12}$, J. J. L{\'o}pez-Moreno$^{17}$, 
F. Marzari$^{4}$, X. Shi$^{3}$, C. Tubiana$^{3}$ 
%%%%% Contributors %%%%%%%%
, B. Gaskell$^{22}$, F. La Forgia$^{4}$ , A. Lucchetti$^{13}$,S. Mottola$^{21}$, M. Pajola$^{13}$, 
F. Preusker$^{21}$, and F. Scholten$^{21}$ 
\end{normalsize}
}%
\vspace{0.5cm}\\
\parbox{\textwidth}{
\begin{normalsize}
$^{1}$LESIA, Observatoire de Paris, PSL Research University, CNRS, Univ. Paris Diderot, Sorbonne Paris Cité, Sorbonne Université, 5 Place J. Janssen, Meudon Cedex 92195, France\\
$^{2}$Center of Studies and Activities for Space (CISAS) G. Colombo, University of Padova, Via Venezia 15, 35131 Padova, Italy\\
$^{3}$Max-Planck-Institut für Sonnensystemforschung, Justus-von-Liebig-Weg, 3, 37077, Goettingen, Germany\\
$^{4}$University of Padova, Department of Physics and Astronomy “Galileo Galilei”, Via Marzolo 8, 35131 Padova, Italy\\
$^{5}$University of Padova, Center of Studies and Activities for Space (CISAS) “G. Colombo”, Via Venezia 15, 35131 Padova, Italy\\
$^{6}$CNR-IFN UOS Padova LUXOR, Via Trasea, 7, 35131 Padova, Italy\\
$^{7}$LATMOS, CNRS/UVSQ/IPSL, 11 boulevard d'Alembert,78280, Guyancourt, France\\
$^{8}$Centro de Astrobiologia, CSIC-INTA, 28850 Torrejon de Ardoz, Madrid, Spain\\
$^{9}$International Space Science Institute, Hallerstrasse 6, 3012 Bern, Switzerland\\
$^{10}$Science Support Office, European Space Research and Technology Centre/ESA, Keplerlaan 1, Postbus 299, 2201 AZ Noordwijk ZH, The Netherlands\\
$^{11}$Jet Propulsion Laboratory, M/S 183-401, 4800 Oak Grove Drive, Pasadena, CA 91109, USA\\
$^{12}$Auburn University, Physics Department, 206 Allison Laboratory, Auburn, AL 36849, USA\\
$^{13}$INAF, Astronomical Observatory of Padova, Vicolo dell'Osservatorio 5, 35122 Padova, Italy\\
$^{14}$University of Padova, Department of Industrial Engineering, Via Venezia 1, 35131 Padova, Italy\\
$^{15}$University of Trento, Faculty of Engineering, Via Mesiano 77, 38121 Trento, Italy\\
$^{16}$INAF Astronomical Observatory of Trieste, Via Tiepolo 11, 34143 Trieste, Italy\\
$^{17}$Instituto  de Astrof\'{i}sica de Andaluc\'{i}a (CSIC), c/ Glorieta de la Astronomia s/n, 18008 Granada, Spain\\
$^{18}$Graduate Institute of Astronomy, National Central University, 300 Chung-Da Rd, Chung-Li 32054, Taiwan\\
$^{19}$Space Science Institute, Macau University of Science and Technology, Avenida Wai Long, Taipa, Macau\\
$^{20}$Institut f\"ur Geophysik und extraterrestrische Physik, Technische Universit\"at Braunschweig, Mendelssohnstr. 3, 38106 Braunschweig, Germany\\
$^{21}$Deutsches Zentrum für Luft- und Raumfahrt (DLR), Institut für Planetenforschung, Asteroiden und Kometen, Rutherfordstraße 2, 12489, Berlin, Germany\\
$^{22}$Planetary Science Institute, 1700 East Fort Lowell, Suite 106, Tucson, AZ, 85719, USA
\end{normalsize}
}
\end{center}
}] % end of single column
\newpage
\twocolumn[{%
%%%%%%%%%%%%%%%%%%% ABSTRACT %%%%%%%%%%%%%%%%%%%%%
\section*{Abstract}
Context:\\
From August 2014 to September 2016, the Rosetta spacecraft followed comet 
67P/Churyumov-Gerasimenko along its orbit. After the comet passed 
perihelion, Rosetta performed a flyby manoeuvre over the Imhotep-Khepry 
transition in April 2016. The OSIRIS/Narrow-Angle-Camera (NAC) acquired 
112 observations with mainly three broadband filters (centered at 480, 
649, and 743nm) at a resolution of up to 0.53 m/px and for phase angles 
between 0.095\dgr and 62\dg.\newline\\
Aims:\\
  We have investigated the morphological and spectrophotometrical properties 
  of this area using the OSIRIS/NAC high-resolution observations.\newline\\
Methods:
  We assembled the observations into coregistered color cubes. Using a 3D 
  shape model, we produced the illumination conditions and georeference 
  for each observation. We mapped the observations of the transition to 
  investigate its geomorphology. Observations were photometrically 
  corrected using the Lommel-Seeliger disk law. Spectrophotometric analyses 
  were performed on the coregistered color cubes. These data were used to 
  estimate the local phase reddening.\newline\\
Results:\\
  The Imhotep-Khepry transition hosts numerous and varied types of terrains 
  and features. We observe an association between a feature's nature, its 
  reflectance, and its spectral slopes. Fine material deposits exhibit an 
  average reflectance and spectral slope, while terrains with diamictons, 
  consolidated material, degraded outcrops, or features such as somber 
  boulders, present a lower-than-average reflectance and higher-than-average 
  spectral slope. Bright surfaces present here a spectral behavior 
  consistent with terrains enriched in water-ice. We find a phase -reddening 
  slope of 0.064$\pm$0.001\%/100nm/\dgr at 2.7 au outbound, similarly to the 
  one obtained at 2.3 au inbound during the February 2015 flyby.\newline\\
Conclusions:\\
  Identified as the source region of multiple jets and a host of water-ice 
  material, the Imhotep-Khepry transition appeared in April 2016, close to 
  the frost line, to further harbor several potential locations with exposed 
  water-ice  material among its numerous different morphological terrain units.%
  \newline\\
Keywords: comets: individual: 67P/Churyumov-Gerasimenko, space vehicles: %
  ROSETTA, space vehicles: instruments: OSIRIS, methods: data analysis, %
  techniques: image processing\newline\\
}] % end of single column

%%%%%%%%%%%%%%%%% BODY OF PAPER %%%%%%%%%%%%%%%%%%
%Section Introduction
\section{Introduction}
As part of the HORIZON 2000 perspective, the ROSETTA mission has been the 
European Space Agency's cornerstone for the study of the small bodies of the 
solar system \citep{Bar-Nun_1993}. For 26 months, the Rosetta spacecraft 
followed comet 67P/Churyumov-Gerasimenko (67P) along its orbit from $\sim$4.3 au 
inbound to perihelion to $\sim$3.8 au outbound. During this period, its 
instruments extensively characterized under different observations 
conditions the nucleus and the inner coma. After dropping the Phil\ae $ $ probe, 
which performed measurements directly on the surface of the nucleus, the spacecraft 
instruments notably monitored the nucleus for changes as the comet approached, 
went through and moved away from its perihelion (reached on 13 August 
2015).\\
In particular, during these 26 months, the OSIRIS instrument, which is the Rosetta 
scientific imaging system \citep{Keller_2007}, acquired a vast number of
observations of the comet in the 200-1000 nm wavelength domain. Most notably, 
during low-altitude flyby manoeuvres over the nucleus, performed in February 
2015 and April 2016, the OSIRIS instrument imaged the nucleus surface at 
different wavelengths with a sub-meter spatial resolution 
(\citealp{Feller_2016} and \citealp{Hasselmann_2017}).\newline\\
The ROSETTA mission and the OSIRIS instrument have especially shown that the 
nucleus surface is exceedingly dark (p$_{v,649nm}$ $\sim$6.7\%), its visible 
spectrum does not exhibit absorption bands, and its reflectivity notably 
increases with the wavelength, that is, it presents a red spectral behavior 
(e.g., \citealp{Sierks_2015, Fornasier_2015}). 
Furthermore, the two lobes of the nucleus present the same range of 
morphologic, spectroscopic, spectrophotometric and photometric properties overall, 
although some subtle differences can be observed at the centimeter to hectometer 
scale in terms of colors, spectra, and composition (e.g., \citealp{Capaccioni_2015, 
Filacchione_2016a, Fornasier_2015, Pommerol_2015, Poulet_2016}, or 
\citealp{Fornasier_2016}).\\
In the appraisal of the pre-perihelion data gathered by the Rosetta spectrometer 
VIRTIS \cite{Coradini_2007} and \cite{Quirico_2016} indicated that the spectrum and 
low albedo of the average nucleus surface can be accounted for as a mixture of 
opaque minerals with dark refractory polyaromatic carbonaceous components 
bearing methyl, alcohol, ammonium, and ester groups.\\
While assessing the nature of the nucleus surface and investigations for
relevant surface analogs are still an ongoing subject of research (e.g., 
\citealp{Jost_2017a, Jost_2017} and \citealp{Rousseau_2017} and references therein), the 
examination of the remaining parts of the trove of images acquired by the OSIRIS 
instrument is also underway. In this study, we present the results of the 
spectrophotometric analysis from some of the most striking features of the 
region observed during the April 2016 flyby.\\
In the next section, we present the observational dataset, the data reduction 
procedures, and the methods used in this analysis, before we briefly present 
the morphological properties of the flyby area in section 3. In section 4 we 
present the results of this spectrophotometric analyses before we discuss our 
findings in section 5.
\section{OSIRIS/NAC observations of the April 2016 flyby}
\label{sec:observations}
\begin{figure*}
 \begin{minipage}[t]{0.525\linewidth}
   \includegraphics[width=\textwidth]{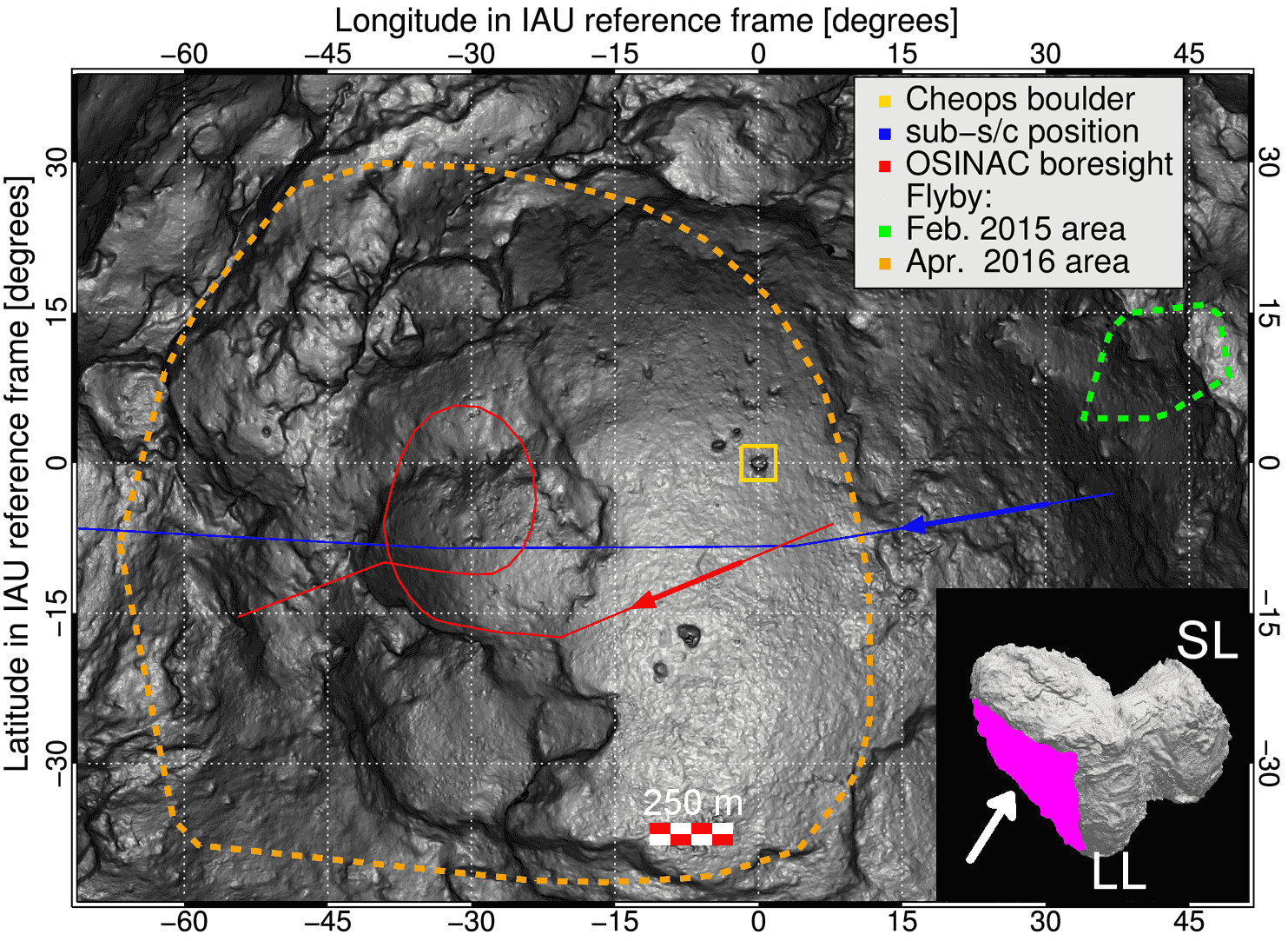}
 \end{minipage}\hspace{1cm}
 \begin{minipage}[t]{0.395\linewidth}
  \begin{center}
  \includegraphics[width=0.95\textwidth]{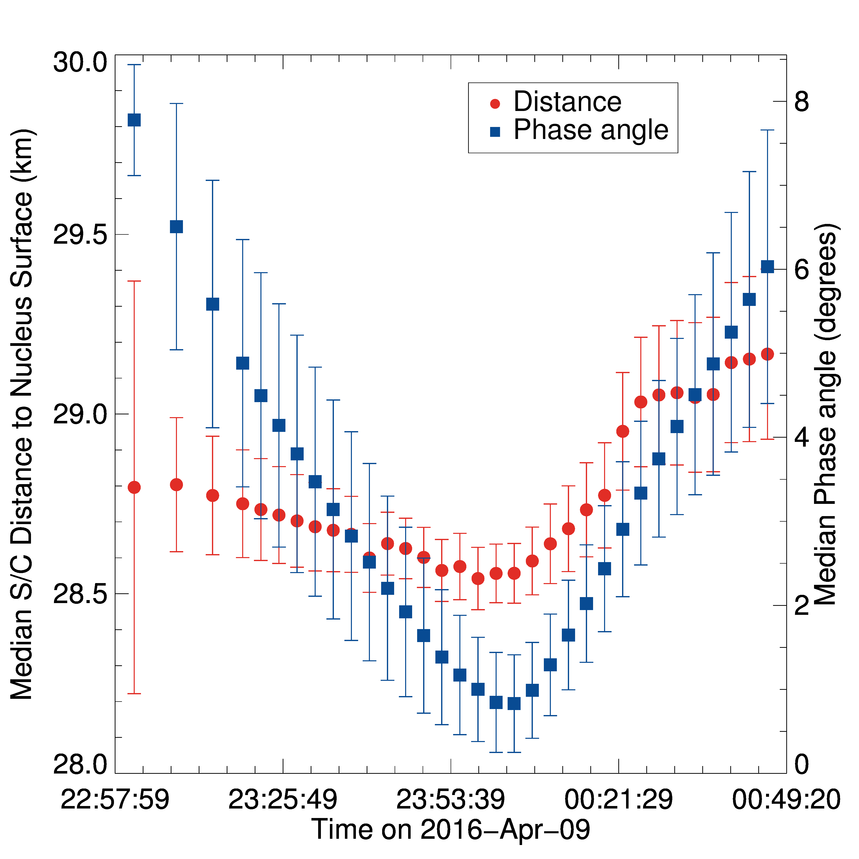}
  \end{center}
 \end{minipage}
 \caption{Context of 9-10 April 2016. Left (inset): Position of the
    Imhotep depression (in pink) with respect to the comet large lobe (LL) and
    small lobe (SL). (Main) Cylindrical projection centered on the Cheops
    boulder (gold square) of the Imhotep region, see main text for more details.
    Right: Curves of the median distance to the surface and median phase
    angle around the moment of closest approach during the flyby. The
    uncertainties correspond to the standard deviation of the values across each
    observation.}
 \label{fig: context flyby}
\end{figure*}

\textit{\textup{The OSIRIS/NAC instrument.}} The scientific imaging system on 
board the Rosetta spacecraft, OSIRIS, comprised two cameras: the Narrow-Angle 
Camera (NAC) and  the  Wide-Angle Camera (WAC). The NAC had a 2048x2048 px CCD 
array, each pixel being a square with an 13.5 $\mu$m edge. The optical 
system associated with the NAC gave it a field of view of 2.35\dg x2.35\dgr 
and an angular resolution of 18.6 micro-radians per px ($\mu$rad / px). The NAC 
also comprised a set of 12 broadband filters optimized for the study of the 
nucleus mineralogy in the 250-1000 nm wavelength domain. A selection of the 
NAC filters relevant to this study is listed in Table \ref{tbl: OSIRIS filters}. 
For a detailed description of the instrument specifications and hardware, we 
refer to \citet{Keller_2007}.\newline\\
\begin{table}
 \centering
 \begin{tabular}{|c|c|c||c|c|c|}
  \hline
  \hline
  Filter & $\lambda_{c}$ & $\Delta\lambda$ & Filter & $\lambda_{c}$ & $\Delta\lambda$ \\
  $ $ &  [nm]    & [nm] & $ $ &  [nm]    & [nm] \\  
  \hline
  F15 & 269.3 & 53.6 & F28 & 743.7 & 64.1\\
  F16 & 360.0 & 51.1 & F51 & 805.3 & 40.5\\
  F24 & 480.7 & 74.9 & F41 & 882.1 & 65.9\\
  F23 & 535.7 & 62.4 & F61 & 931.9 & 34.9\\
  F22 & 649.2 & 84.5 & F71 & 989.3 & 38.2\\
  F27 & 701.2 & 22.1 & $ $ & $   $ & $  $\\ 
  \hline
 \end{tabular}
 \caption{ \label{tbl: OSIRIS filters} OSIRIS/NAC filters: the table lists 
    the filters of the OSIRIS/NAC instrument with their associated central 
    wavelength and bandwidth.}
\end{table}
\textit{\textup{Context of the observations.}} At the time of the April 2016 flyby,
comet 67P was outbound from perihelion and its heliocentric distance
increased from 2.76 au to 2.78 au between 9 and 10 April.
In order to perform this particular flyby manoeuvre, the Rosetta spacecraft
was moved more than 950 km away from the comet two weeks before the manoeuvre 
was executed. It was then progressively approached again, and flown above the 
transition area between the Imhotep and Khepry morphological regions at less 
than 30 km from the nucleus surface, at the moment of closest approach.\\
The position of this transition on the nucleus, the area common to all OSIRIS 
images acquired during this manoeuver as well as that from the February 2015 flyby, 
are shown in the left panel of fig. \ref{fig: context flyby}. The temporal evolutions of 
the median distance between spacecraft and the imaged surfaces as well as the 
median phase angle for each observation taken during the flyby are plotted 
in the right panel of fig. \ref{fig: context flyby}.\\
The diagram displayed in the right panel of fig. \ref{fig: context flyby} corresponds to 
the evolution of the median phase angle (blue squares) within an observation's 
frame, around the time of closest approach.\newline\\
The left panel of Fig. \ref{fig: context flyby} was produced using the “SPG SHAP7 v1.0” shape
model with five millions facets and a horizontal spacing of about 2 meters 
\citep{Preusker_2017}. In this figure, which we only produced for illustrative purposes, 
the brighter a facet, the smaller the angle between its normal and the 
direction from the nucleus center of mass to its barycenter. This means that the 
relative brightness of a facet in the left panel of fig. \ref{fig: context flyby} is 
indicative of the local tilt. Additionally,  the longitudes and latitudes are 
given here relative to the Cheops boulder (gold square) in order to 
properly accommodate the common area to the observations taken during the 14 
February 2015 flyby (defined by the green dashed line), which lay just around 
the -180\dg/+180\dgr border in the Cheops reference frame \citep{Preusker_2015}. 
In this frame, the coordinates of the Cheops boulder are +142.35\dgr E, 
-0.28\dgr S.\newline 
Additionally, in the left panel of fig. \ref{fig: context flyby}, the area common to all 
NAC observations taken during the flyby manoeuvre is delimited by the dashed 
orange line, while the sub-spacecraft position is marked by the continuous 
blue line. In this figure, the projection of the boresight of the OSIRIS NAC 
on the surface of the comet is indicated by the continuous red line.\\
Alongside the boresight of the OSIRIS/NAC, around the moment of closest
approach, the median distance between the spacecraft and the nucleus surface
varied between 28.6 and 29.2 km, as illustrated by the red dots in the right panel of Fig.
\ref{fig: context flyby}. Hence, OSIRIS/NAC resolved the observed 
terrain with a resolution of about 0.53 m/px.\newline\\
\textup{}Observations. We list in Table \ref{tab: Flyby 2016 NAC images}
the OSIRIS/NAC observation sequences used in this study, as well their
corresponding NAC filter image combination.  We also list the median distance
of the spacecraft to the surface elements, the corresponding spatial resolution
for the NAC, the median phase angle, and the phase angle range as well as the
position in longitude and latitude of the NAC boresight.\\
These observations were acquired around the moment of closest approach to the
nucleus and either on an inbound or outbound trajectory from the moment of
closest approach, which was reached during the first minutes after midnight on 10 April 2016.\\
For the analyses of this study, we chose to consider not only the
observations acquired just around the moment of closest approach as a sequence
of 3 NAC filter images (those centered at 480, 649, and 743 nm), but also two 
other sets of observations acquired shortly before and after the moment of 
closest approach using all of the 11 NAC filters (thus spanning the 269.3-989.3 
nm wavelength domain), which also imaged the area in question under notably 
different illumination conditions (see the first and last lines in Table 
\ref{tab: Flyby 2016 NAC images}).\\
When these two additional sets of observations were included in our analysis, we 
found no evidence of morphological and structural differences that might indicate a major activity event in this area around the moment of closest approach.%
\newline
As described in detail in \citet{Kueppers_2007} and \citet{Tubiana_2015}, when 
they were uploaded to Earth, these OSIRIS observations were reduced using the 
standard OSIRIS pipeline (the 1.0.0.34 version of the “OsiCalliope” software). 
In short, all these images were uncompressed, calibrated, radiometrically 
corrected, and converted from DN/s to radiance factor\footnote{The radiance 
factor (or RADF, \citealp{Hapke_1993}) is also denoted hereafter as I/F. We 
recall that the radiance factor and the reflectance factor (or REFF, 
\citealp{Hapke_1993}), often used in laboratory experiments and for instance in 
\citet{Jost_2017a}, are related as follows: REFF = RADF/cos(i), where i is 
the solar incidence angle.} , and they were also corrected for geometric distortion.%
\newline\\
Similarly to \citet{Fornasier_2015,Feller_2016} and \citet{Hasselmann_2017}, 
two further steps were added before the analyses: (1) we computed and retrieved 
the illumination conditions of the surface elements intercepted by each pixel 
of each image of the dataset, (2) we coregistered each image from each 
observation sequence to its corresponding F22 image, so as to force a 
correspondence between the pixels of each and every image of an observation 
sequence with the intercepted surface elements.\\
These two steps were achieved using software developed within the OSIRIS
team. The first was assembled based partially on
\citet{Gaskell_2008} and \citet{Jorda_2010}. The latter step was written based 
on \citet{VanderWalt_2014} as a two-part process in which first, 
image features are detected with the “Oriented FAST and Rotated BRIEF” (ORB) 
algorithm (\cite{Rublee_2011} or through an image segmentation and the use of 
an optical-flow algorithm (as based on \citealp{Lucas_1981, 
Shi_1994} and \citealp{Bugeau_2009}), and then, when the features were matched, each image was 
coregistered to the reference image using a homographic transformation.%
\newline\\
In order to produce mappings of the illumination conditions, we used the 
“SPG SHAP7 v1.0” shape model of the nucleus, as detailed before. 
Furthermore, since at the time of these observations, the comet was only at 2.7 
au, the Sun cannot be considered as a point-like source here. As noted in 
\cite{Shkuratov_2011},  the trigonometric relation expressing the angular 
diameter of Sun across the nucleus surface is arcsin(R$_{\odot}$/r), where 
R$_{\odot}$ is the radius of the Sun's photosphere and r is the distance of the 
observed object from the Sun.\\
Hence, we can only sample phase angles greater than 0.095\dg. Therefore the phase 
angle ultimately ranges in our dataset from 0.095\dgr to 61.7\dg. In Table 
\ref{tab: Flyby 2016 NAC images}, phase angle ranges that have been truncated 
thus are denoted by a star.
%
%Section Morphology
\section{Morphology of the flyby area}
\label{sec:3 morphology}
\begin{figure*}
 \begin{minipage}[t]{0.47\linewidth}
  \begin{center}
   \includegraphics[width=\linewidth]{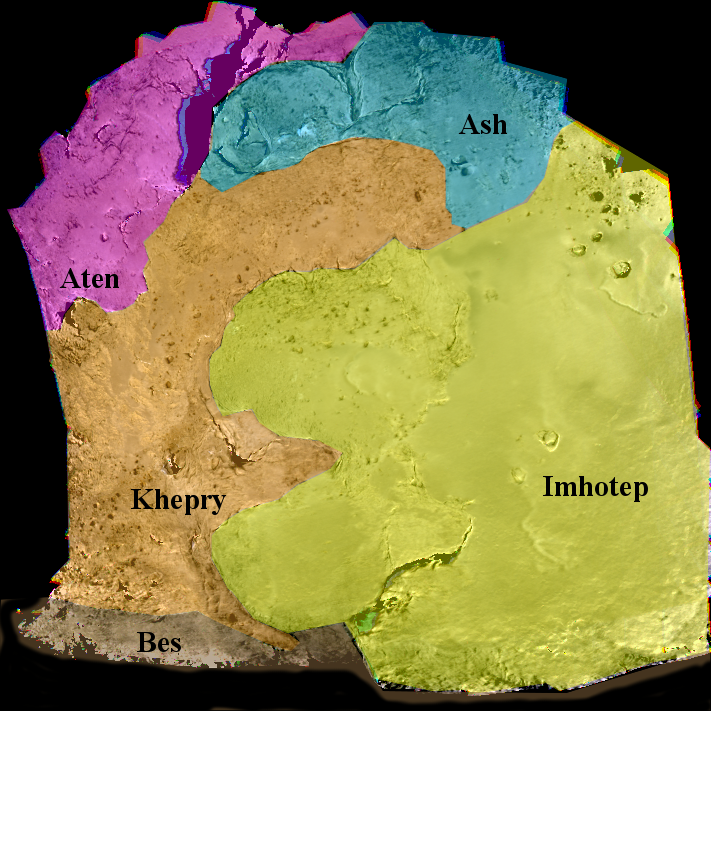}
  \end{center}
 \end{minipage}
 \begin{minipage}[t]{0.47\linewidth}
  \begin{center}
   \includegraphics[width=\linewidth]{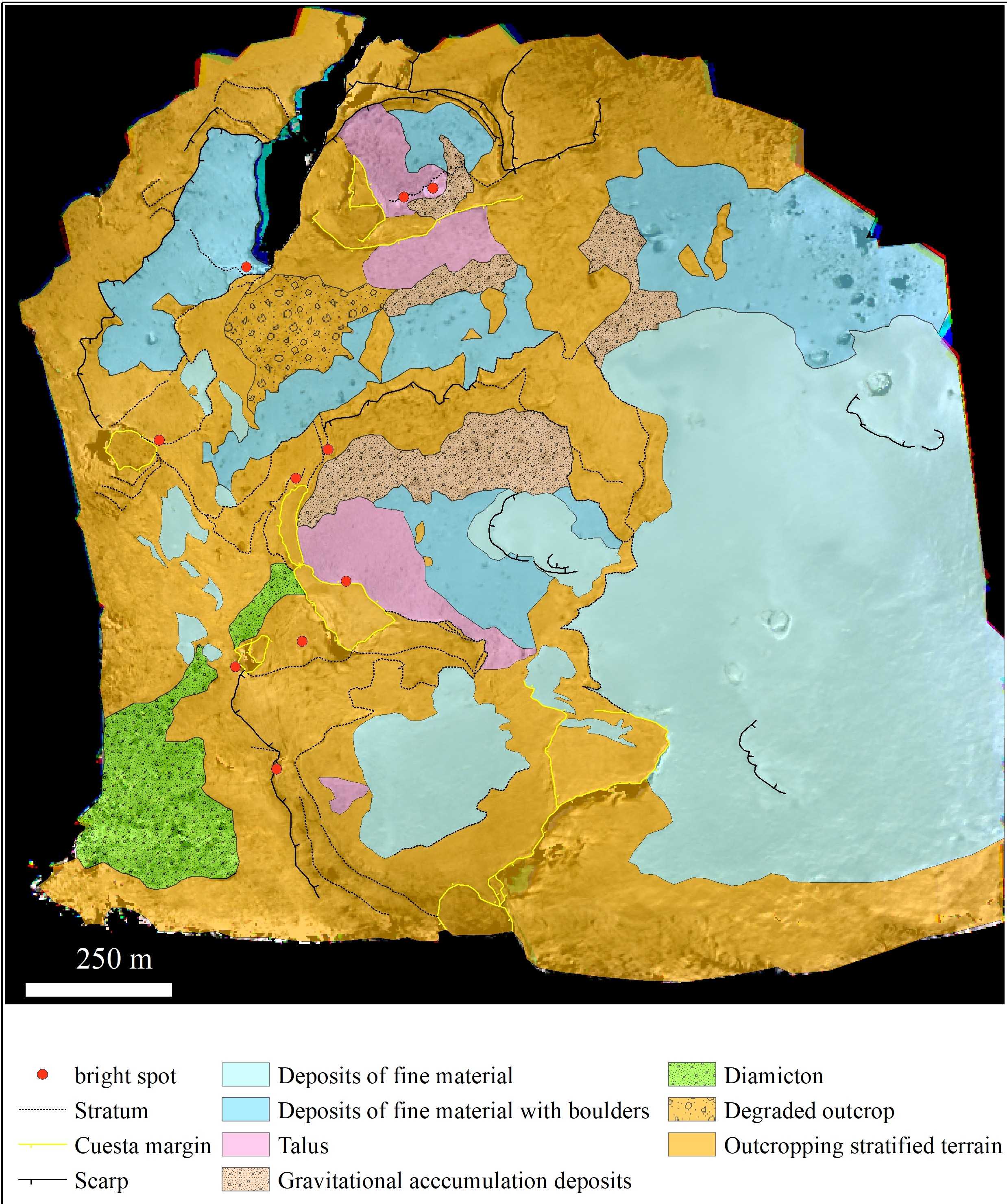}
  \end{center}
 \end{minipage}
 \caption{Left: RGB mapping of the flyby area overlaid with the limits of the
    corresponding morphological regions (see \protect\citealp{El-Maarry_2015}). 
    Right: Geomorphological mapping of the transition area between Khepry and 
    Imhotep. The peculiarities of the bright spots are discussed in the
    following sections.}
 \label{fig: geomorpho}
\end{figure*}

Throughout the ROSETTA mission, the nucleus of comet 67P has been shown to 
comprise a variety and a complexity of aspect and morphological structures, as  
discussed at length in \citet{Thomas_2015, El-Maarry_2015, Auger_2015,
Massironi_2015} and \citet{El-Maarry_2016}. The depression that is the Imhotep 
physiographical region is surrounded by the Apis, Ash, Bes, and Khepry regions.\\
As the spacecraft flew over this part of the comet (see the left panel of Fig. 
\ref{fig: context flyby}), OSIRIS/NAC acquired observations of the Imhotep 
depression and of the nearby regions Ash,  Aten, Bes, and Khepry (see the 
right panel of fig. \ref{fig: geomorpho}). Given the path of the spacecraft in 
these observations, we hereafter refer to the area of the April 2016 flyby as 
the Khepry-Imhotep transition.\newline\\
\begin{figure*}
 \begin{minipage}[t]{0.49\linewidth}
   \begin{center}
   \includegraphics[width=0.85\linewidth]{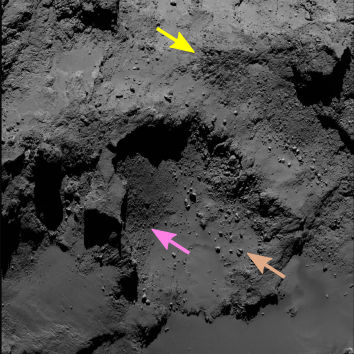}
   \end{center}
 \end{minipage}
 \begin{minipage}[t]{0.49\linewidth}
   \begin{center}
   \includegraphics[width=0.85\linewidth]{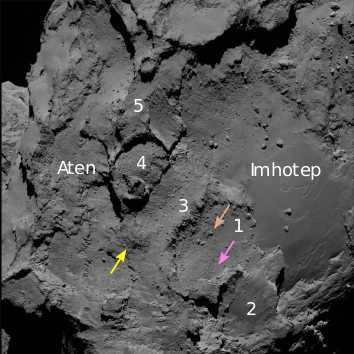}
   \end{center}
 \end{minipage}
 \caption{OSIRIS NAC observations of the Khepry-Imhotep transition region. 
    Left (NAC 2016-06-09T08.47.10.761Z F22) arrows point 
    toward degraded outcrops, talus, and gravitational deposits at the boundary 
    between the Aten, Imhotep, and Khepry regions.\\ 
    Right (NAC 2014-09-02T15.44.22.578Z F22): Between niches, consolidated 
    materials form an ordered staircase structure, which from the deeper terrace 
    labeled 1 rises up to the highest terrace labeled 5.}
 \label{fig: additional morpho}
\end{figure*}
As detailed in \citet{El-Maarry_2015} and \citet{El-Maarry_2016}, these 
regions all present different aspects and structure, but also common 
characteristics. While Ash, Aten, Bes, and Khepry present common features such 
as exposed consolidated material in the form of terraced and layered units 
and smooth deposits, Ash has been shown to be an area of airfall material 
deposit \citep{Thomas_2015a}, and Bes hosts fractures and a pit.\\
Since the unconsolidated terrains of Imhotep are located in wide flat areas that 
correspond to the local gravitational lows, \citet{Auger_2015} defined the 
region as a real accumulation basin for boulders and fine materials. In 
addition, along the side of the region that is being investigated and just 
before the perihelion passage, they observed that consolidated materials 
displayed bright patches despite their long-lasting exposure to illumination. 
This regeneration of bright patches on scarps has been associated with a 
progressive retreat driven by sublimation.\newline\\
In the Khepry-Imhotep transition, as illustrated in the right panel of fig. 
\ref{fig: additional morpho}, most of the surface terrains consist of 
consolidated materials that form terraces and niches because the Bes, Khepry, 
and Ash regions decline either toward the Imhotep or the Aten depressions 
\citep{Giacomini_2016} and \citep{Lee_2016}. Imhotep in particular is 
indeed at a deeper structural level with respect to all these regions 
\citep{Penasa_2017}, and on this specific side, is mostly filled by deposits of 
fine unconsolidated materials comprising low-density boulder fields or isolated 
megaclasts \citep{Auger_2015} and nucleating scarps. Outcropping layered 
terrains that surround depressions and niches form an ordered staircase 
structure of terraces (see the right panel of fig. \ref{fig: additional morpho}).\\
The localized erosion of these terraces provides coarse surfaces (i.e., degraded 
outcrops, see the area indicated to by the yellow arrow in the right panel of fig. 
\ref{fig: additional morpho}) or more evolved clast fields (i.e., 
diamictons) on their flat top, or clast deposits at the cliff base. The latter 
can be ascribed to a specific source and are distinguished in talus eposits, 
which consist of well-sorted clasts and gravitational deposits, which in 
constra imply well-graded unconsolidated materials supported by fine materials 
(see the left panel of fig. \ref{fig: additional morpho}, the pink and brown 
arrows, respectively).\\
Terraces are partly covered by shallow deposits of fine material (i.e., 
unresolved regolith), which could likely be either the result of in-situ 
erosion or of airfall deposits, whereas their overhangs and steep ramps host 
most of the observed bright patches (bright spots in the right panel of fig. \ref{fig: geomorpho}).
%
%Section Spectrophotometric results
\section{Global and local spectrophotometry}
In the following section, we present the results of the spectrophotometric 
analysis of the region and of particular surface elements from the OSIRIS/NAC 
observations that were taken just around the moment of closest approach and were acquired 
using the filters centered at 480, 649, and 743 nm.
\subsection{Global spectrophotometry}
\begin{figure*}
 \begin{minipage}[t]{\linewidth}
  \begin{center}
   \includegraphics[height=0.45\linewidth]{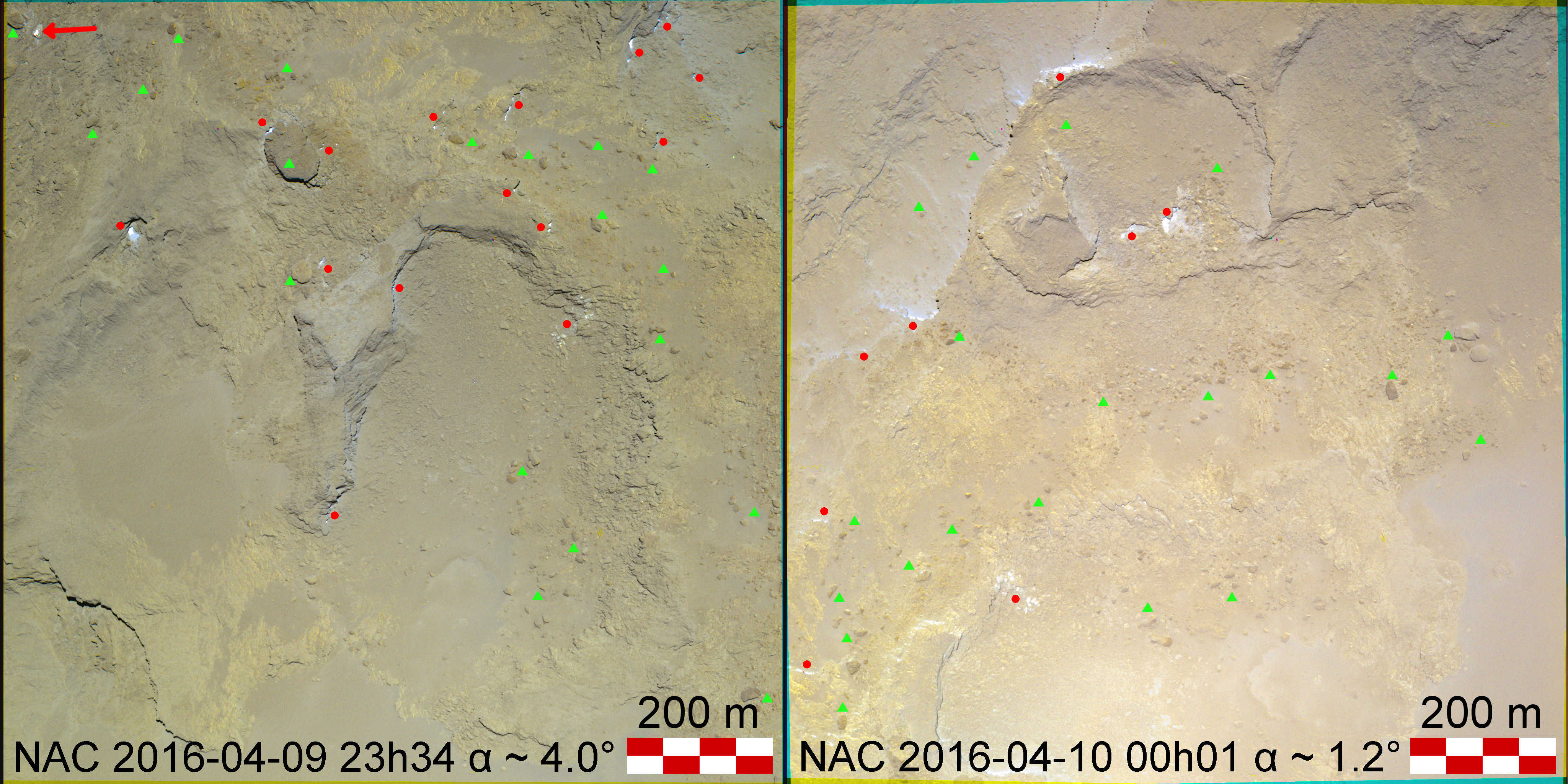}
  \end{center}
 \end{minipage}
 \begin{minipage}[b]{\linewidth}
  \begin{center}
   \includegraphics[height=0.5\linewidth]{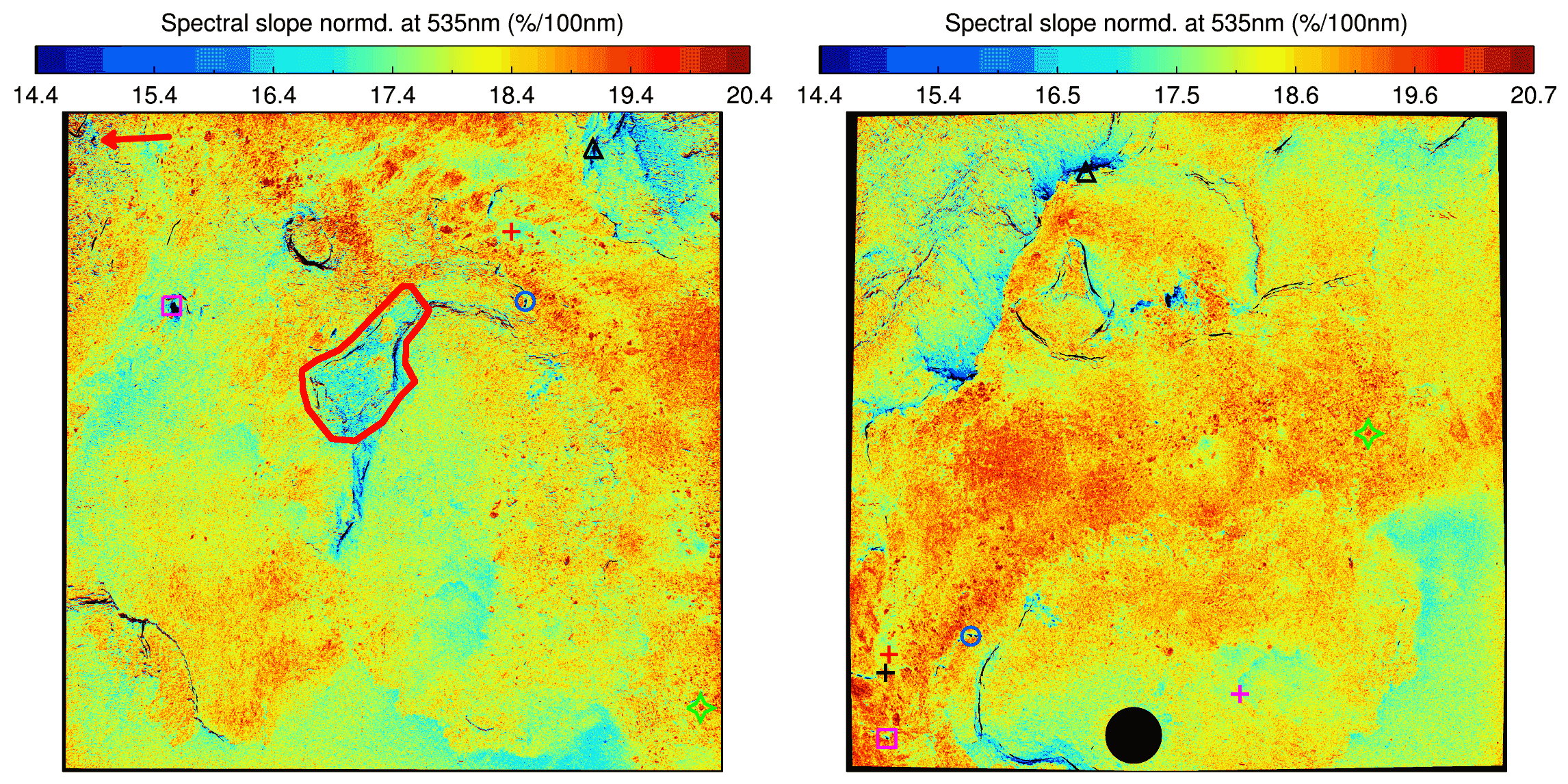}
  \end{center}
 \end{minipage}
 \caption{Top: Panel of RGBs produced from 480nm, 639nm, and 743nm NAC filter
    images taken during the flyby. For clarity, the 2016-04-10T00h01 image has
    been rotated anticlockwise by 90\dgr. The positions of the bright surfaces
    discussed in the main text are denoted by red circles, while locations of
    groups of somber boulders are marked by green triangles. Bottom: Panel of
    the spectral slopes in the 535--743 nm range for the NAC images presented
    in fig. \ref{fig: Flyby 2016 I/F}. The black circle in the right mapping 
    encloses the surface elements for which the phase angle is lower than
    0.095\dg. The symbols used in each image correspond to the surface
    elements investigated in fig. \ref{fig: Flyby 2016 3 filters spectro}.
    See main text for details regarding the black and pink crosses and
for    the central cuesta structure encircled here in red. The red arrow points
    to the position of bright spot 44 in
    \protect\citet{Deshapriya_2018}.}
 \label{fig: Flyby 2016 panels}
\end{figure*}

\begin{figure*}
  \begin{center}
   \includegraphics[height=0.6\linewidth]{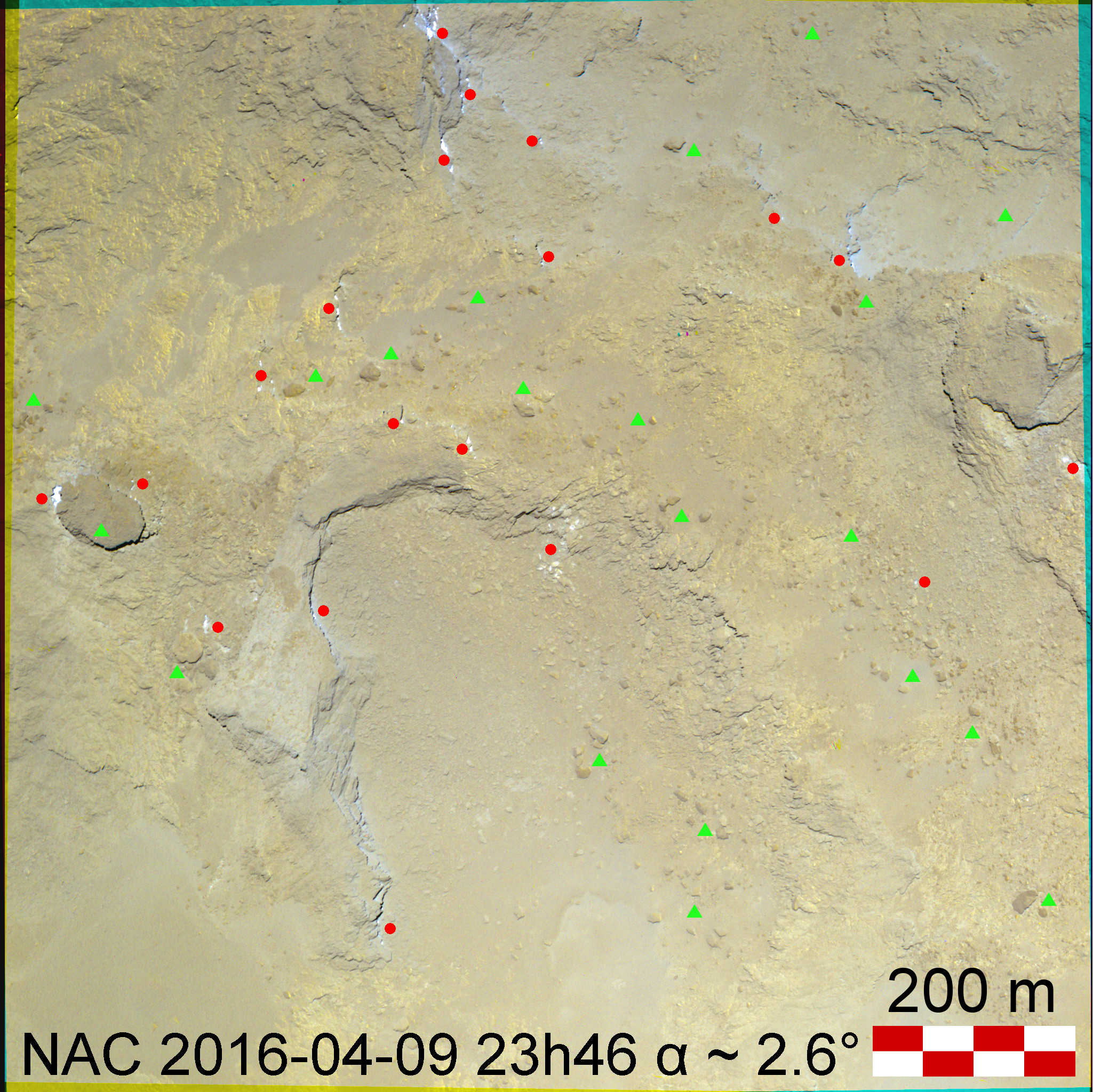}
  \end{center}
  \caption{RGB from NAC filter images acquired at 23h46, before the moment of 
     closest approach (see fig. \ref{fig: Flyby 2016 panels}). The red circles 
     point to the location of bright surfaces, while the green triangles 
     indicate groups of somber boulders. This figure encompasses all of the features 
     investigated in this paper.}
  \label{fig: Flyby 2016 central RGB}
\end{figure*}

\begin{figure}[!h]
  \centering
  \includegraphics[height=0.9\linewidth]{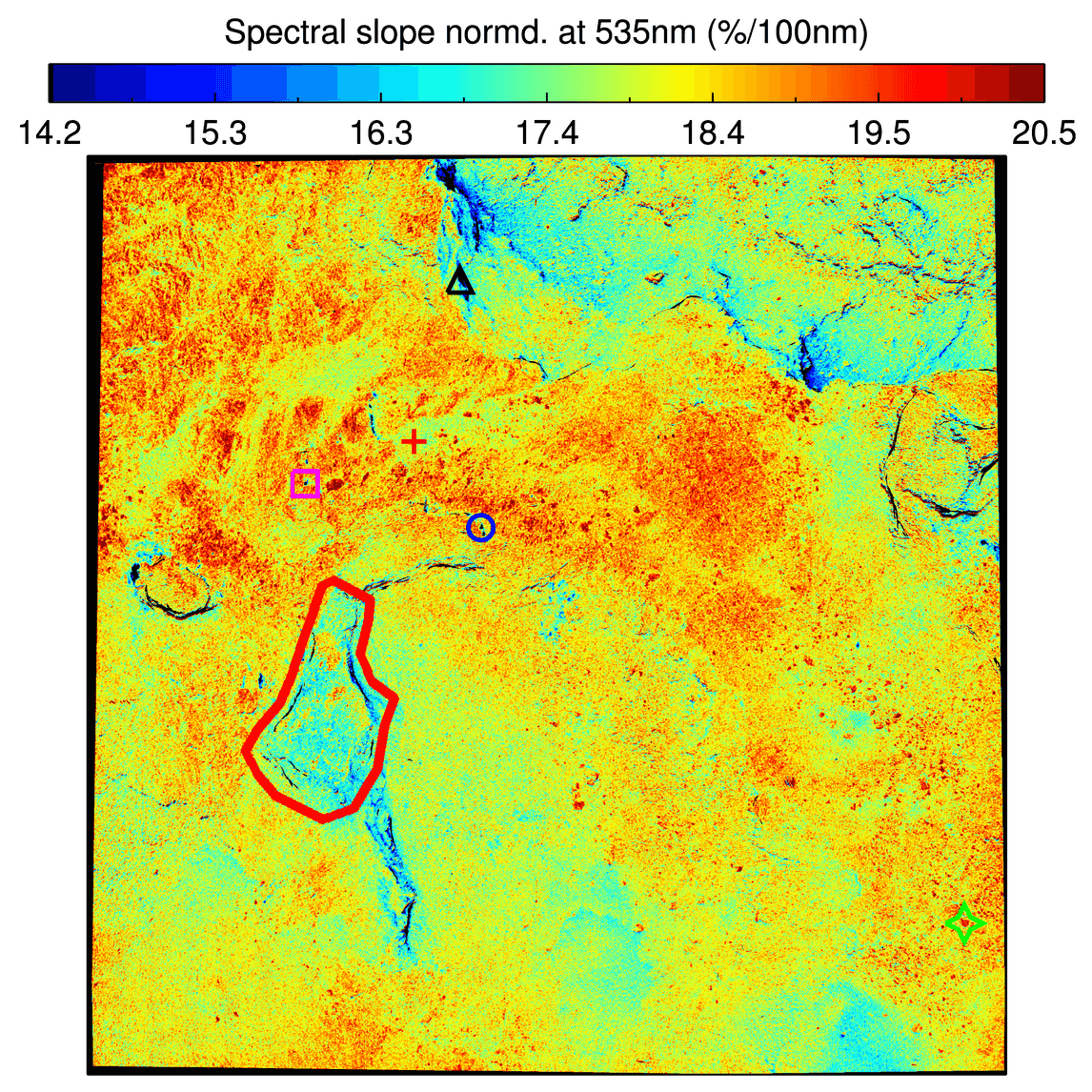}
  \caption{\label{fig: Flyby 2016 central slp}
     Spectral slope in the 535--743 nm range associated with fig. 
     \ref{fig: Flyby 2016 central RGB}. The symbols used here and in the bottom 
     panel of fig. \ref{fig: Flyby 2016 panels} refer to the same surface 
     elements. See fig. \ref{fig: Flyby 2016 3 filters spectro} for the 
     corresponding measures. The red polygon defines the outcropping of 
     spectrally blue consolidated material that is also marked in fig. 
     \ref{fig: Flyby 2016 panels} and is discussed in the main text.}
\end{figure}

As shown in the right panel of fig. \ref{fig: context flyby} and in Table 
\ref{tab: Flyby 2016 NAC images}, most of the NAC images of this dataset were 
acquired with a pixel scale of $\sim$0.53 m/px. In these observations,  
the phase angle densely samples the range between 0.1\dgr and 10\dg. Most of the shadows are therefore absent in these panels, and the contents of the niches, 
which are usually shadowed, are visible.\\
False-color images (hereafter RGBs), assembled from the NAC observations 
that wer acquired during the flyby area, are presented in the top panels of figs. 
\ref{fig: Flyby 2016 panels} and \ref{fig: Flyby 2016 central RGB}. 
The spectral slope mappings (computed in the 535-743 nm range and normalized 
at 535 nm) corresponding to the RGB figures are shown in the bottom panels of figs. 
\ref{fig: Flyby 2016 panels} and \ref{fig: Flyby 2016 central slp}. 
As described in \citet{Fornasier_2015}, these spectral slopes were computed 
using the formula in Eq. \ref{Eq:Spectral slope definition},\begin{equation}
S [\%/100nm] = \frac{R_{743nm}-R_{535nm}}{R_{535nm}}\cdot \frac{10^{4}}{ (743.7 [nm] - 535.7 [nm])}
\label{Eq:Spectral slope definition}
.\end{equation}
As in \cite{Feller_2016}, when absent, the radiance factor mappings at 535 nm 
were estimated in this study by interpolating between those observed at 480 
nm and those at 649 nm. The resulting differences are discussed in section \ref{sec:section_4_4}.\newline
The spectral normalization was performed with the filter centered at 535 nm as no 
mineralogical bands are expected there, and in order to emulate the 
normalization using the V filter \citep{Bessell_1990} in the small-body 
literature.\newline\\
In the RGB panorama\footnote{Assembled from 2016-02-10 15h20 and 15h28 
OSIRIS/NAC images taken with a 0.9 m/px resolution and at a phase angle of 
$\sim$65\dg, and as such comparable to fig. \ref{fig: spc slp 12h13}.} (fig. 4) 
of \citet{Hasselmann_2017}, the transition area between the Khepry and Imhotep 
regions appeared to only slightly vary in terms of colors and spectral slopes, 
in the considered dataset and in the top panels in figs. \ref{fig: Flyby 2016 panels} and 
\ref{fig: Flyby 2016 central RGB} for instance, given the higher resolution 
and the lower range of phase angles, several additional structures and finer 
variations are evident.\newline\\
The most striking features in this area are the bright blue-white patches 
located either in niches, underneath overhangs, or also among, at the tops of, 
or behind boulders. These features all present spectral slopes that are smaller 
than those of their surroundings. The position of the bright spots we 
investigated is marked by red circles in the top panels of figs. 
\ref{fig: Flyby 2016 panels} and \ref{fig: Flyby 2016 central RGB}. A cursory 
investigation of the surfaces that are located close to overhangs and have a 
spectral slope smaller than 14 \%/100 nm at these low phase angles (see fig. 
\ref{fig: Flyby 2016 central slp}) points to circumferences that vary between 
2 and at least 12 pixels (i.e., between $\sim$ 1 and 6 meters). The widest 
bright surface measured is located below the black upward triangle in the 
bottom right panel of fig. \ref{fig: Flyby 2016 panels}-b. Additionally, within 
the niche indicated by the pink square in the  bottom left panel of fig. 
\ref{fig: Flyby 2016 panels}, the area of the whole bright surface is $\sim$ 
160 m$^{2}$.\newline\\
In the bottom right panels of figs. \ref{fig: Flyby 2016 panels} and 
\ref{fig: Flyby 2016 central slp}, another striking feature is the apparent 
unit of outcropping consolidated material with the inclined top (cuesta) that 
is delimited by the red polygon. The overall spectral behavior of this feature 
is smaller than its surroundings, which means that it might indicate an 
enrichment in water-ice rich material on the whole surface of the outcrop. The 
overall spectral slope of this unit is $\sim$ 17 \%/100 nm at 4\dgr phase 
angle, while its surroundings are at $\sim$ 18 \%/100 nm. This outcrop is the 
only such large unit of consolidated and stratified material in the flyby area 
that shows this behavior.\\
We furthermore observe that the surface of this terrain presents small variations 
of spectral slopes that closely follow the nature of the terrain: in contrast to 
terrains with a smooth aspect or those that lie at the top of the local cliffs, 
terrains presenting a rougher aspect or that have a higher declivity exhibit a 
slightly stronger spectral slope.\newline\\
We also note that the spectral behavior of areas that are covered with fine 
material deposits (see the right panel of fig. \ref{fig: geomorpho}) is smaller 
than that of their surroundings, but only slightly. For instance, in the bottom 
right panel of fig. \ref{fig: Flyby 2016 panels}, the material found at the top 
of the scarp (its location is indicated by the center of the pink cross) has 
a reflectance of 6.81$\pm$0.02\% at 649 nm and a spectral slope of 17.9$\pm$0.1 
\%/100 nm, while the crown of the scarp directly next to it (its location is 
found just below the right arm of the pink cross, 13 pixels, about $\sim$6.9 m 
away from its center) has a corresponding reflectance of 6.54$\pm$0.02\% and 
a spectral slope of 18.2$\pm$0.2 \%/100 nm. This amounts to a mere 4.0\% 
difference of the reflectance and only 1.7\% disparity in spectral slope.\\
Similarly, between the red cross that marks the position of a fine material 
 deposit (also referred to hereafter as UR-01), and the black cross just below 
it, which indicates the top of a consolidated material area whose surface 
is unresolved and apparently smooth, we find that the fine material deposit 
has a reflectance of 6.37$\pm$0.03\% at 649nm (i.e., 3.8\% lower than the 
consolidated material surface) for a spectral slope of 18.2$\pm$0.2 \%/100 nm 
(i.e., 5.5\% lower than the consolidated material).\\
These two examples illustrate the contrast and differences between these 
surface elements of the nucleus and that they can be perceived, even though 
they are weak.\newline\\
Following this correlation between spectral behavior and terrain nature, we 
furthermore observe that in terrains that are covered with diamicton and areas where consolidated 
material emerges, talus and gravitational accumulation deposits are  
particularly distinguishable in the RGBs through their ochre-orange tinge, and 
in the spectral slopes mappings, they have values higher 18 \%/100 nm.\\
A close inspection of the mappings further indicates that their 
variations are associated with a particular morphological feature. Most notably 
among the boulder fields, gravitational accumulation deposits, and outcrops of 
consolidated material, we also observe a widespread presence of large somber boulders 
and megaclasts among the different types of terrains in this area (see the right panel in fig. 
\ref{fig: geomorpho}). The most obvious of the somber features is 
the 60 x 100 m$^2$ sized megaclast visible in the left panel of fig. 
\ref{fig: Flyby 2016 panels} and in fig. \ref{fig: Flyby 2016 central RGB}.\\
Fig. \ref{fig: Flyby 2016 I/F} depicts the radiance factor in the orange filter 
images corresponding to the RGB presented in figure \ref{fig: Flyby 2016 panels}. 
We recall here that these images were photometrically corrected using the 
Lommel-Seeliger disk law (\citealp{Seeliger_1885} and \citealp{Fairbairn_2005}). In the bottom right panels of figs. 
\ref{fig: Flyby 2016 panels} and \ref{fig: Flyby 2016 I/F}, the 
black circle at the bottom of the 2016-04-10T00h01 mappings delimits surface 
elements with phase angles smaller than 0.095\dg, as discussed in 
section \ref{sec:observations}.%\newline\\
\FloatBarrier
\begin{table*}[!h]
  \centering
  \begin{tabular}{|l|c|c|c|}
   \hline
   \hline
   Time (UTC)                         & 23:34          & 23:46           & 00:01                     \\
   \hline
   Median phase angle (degrees)       & 4.0$\pm$1.3    & 3.0$\pm$1.0     & 1.2$\pm$0.6    \\
   Median I/F (at 649nm)              & 5.3$\pm$0.2\%  & 5.8$\pm$0.2\%   & 6.3$\pm$0.30\% \\
   Median S$_{535-743 nm}$ (\%/100nm) & 18.0$\pm$0.6\% & 18.2$\pm$0.6\%  & 18.3$\pm$0.6\% \\
   \hline
   Brightest I/F value (at 649 nm)    & 20$\pm$3\%     & 12.0$\pm$1.3\%  & 14.0$\pm$1.7\% \\
   ID of measure                      &   BF-01*       & BF-08           & BF-06          \\
   Symbol associated                  & red arrow & \color{black}{$\boldsymbol{\triangle}$} & 
                                               \color{blue}{$\boldsymbol{\bigcirc}$}      \\
   S$_{535-743 nm}$ (\%/100 nm)       & 8$\pm$11       & 4$\pm$1         & 5$\pm$3.7      \\ 
   \hline
   Darkest I/F value (at 649 nm)      & 5.1$\pm$0.1\%  & 5.5$\pm$0.1\%   & 5.7$\pm$0.1\%  \\
   ID of measure                      &   SB-05        & SB-05           & SB-05          \\
   Symbol associated                  & - & \huge{\color{green}{$\diamond$}} & -          \\
   S$_{535-743 nm}$ (\%/100 nm)       & 19.5$\pm$0.3   & 19.7$\pm$0.2    & 19.5$\pm$0.3   \\
   \hline
  \end{tabular}
 \caption{\label{tab: Flyby 2016 stats} Details of the median and extreme 
    radiance factor values in the measurements: the symbols are those used in 
    Figs. \ref{fig: Flyby 2016 panels}, \ref{fig: Flyby 2016 I/F}, and 
    \ref{fig: Flyby 2016 3 filters spectro}. The BF-01 measure is marked because 
    this particularly bright surface has also been investigated as the ID-44 
    bright surface in \protect\cite{Deshapriya_2018}; it survived for several 
    months. The positions and the values of other particular measurements are 
    reported in Table \ref{tbl: spectral slopes tbl}.
    }
\end{table*}
We assembled in Table \ref{tab: Flyby 2016 stats} a summary of the median 
and some extreme values of the reflectance at 649 nm (as well as the
corresponding spectral slopes). These latter values correspond to the extrema 
found among the surface elements we investigated, and they are further discussed in the 
next sub-section.\\
In fig. \ref{fig: Flyby 2016 I/F} a progressive overall decrease in
contrast can be perceived as the spacecraft approached the moment of opposition, 
which indicates the observation of the opposition effect (e.g., 
\citealp{Seeliger_1885} and \citealp{Shkuratov_2011}). Nevertheless, many features whose 
reflectances deviate by more than a sigma from the median I/F value are 
distinguishable from their surroundings.\\
This the case, for instance, for bright surfaces, which, as discussed 
previously, are found in niches, at the bottom of underhangs, or at the top 
of boulders. They are easily identifiable in fig. \ref{fig: Flyby 2016 I/F} 
as the white areas. Meanwhile, at the other end of reflectances, 
most of these easily identifiable surface elements are the somber boulders 
discussed previously that we indicate with green triangles in the top panel of figs. 
\ref{fig: Flyby 2016 panels} and \ref{fig: Flyby 2016 central RGB}.%
\newline\\
The values listed in table \ref{tab: Flyby 2016 stats} show that the BF-01 
measure is in particular distinguishable because it is close to four times the median 
reflectance of the observed area and has a standard deviation of 3\%.
This particular highest value measured for the radiance factor corresponds 
to one peculiar bright surface, referred to as BF-01 and as ID-44 in 
\cite{Deshapriya_2018}. It is located in a field of boulders and diamicton 
of the Khepry region, and its position is indicated by a red arrow in the 
upper left corner of figs. \ref{fig: Flyby 2016 panels} and 
\ref{fig: Flyby 2016 I/F}. This particular feature covers part of the boulder 
top. As it is extended well beyond the integration box, the measurement 
we report was performed around the brightest pixel (I/F $\sim$ 24\%). We note 
that the DN/s values of the corresponding pixels are similar to but do not reach 
the saturation levels of the detector. This particular detail was also noted 
in most of the measurements performed over the course of several months that were 
reported in \cite{Deshapriya_2016}. This surface element is further discussed 
in the section on the 11-filter spectrophotometry.\newline 
Nevertheless, this feature is the brightest observed in the region of this 
flyby. For reference, in that same image, when we measure the radiance factor 
of BF-08 (the large bright patch close to an overhang, noted by a black 
triangle), we find a value of 12$\pm$1.3\%/100 nm. As discussed below, the same bright spot was also observed on the next day at 11:50, and still 
had, at a phase angle of $\sim$64\dg, a radiance factor of 3.6$\pm$0.7\%, 
while its surroundings appeared far darker, with a radiance factor of only 
about 1\%.\newline\\
\begin{table*}
 \centering
 \begin{small}
  \begin{tabular}{|c|c|c|c|c|c|c|c|c|}
   \hline
   \hline                                                 
   Feature& UR-00      & BF-02     & BF-05     & BF-06     & BF-08     & BF-010     & SB-05 \\
   \hline                                                 
   23h34  & 17.7$\pm$0.2 & 7.0$\pm$1.5 & 5.0$\pm$5.0 & 7.0$\pm$2.9 & 2.0$\pm$1.6 & $ $         & 19.5$\pm$0.3 \\
   23h46  & 18.0$\pm$0.5 & $ $         & 6.0$\pm$2.0 & 5.0$\pm$4.6 & 4.0$\pm$1.1 & $ $         & 19.7$\pm$0.2 \\
   00h01  & 18.3$\pm$0.2 & $ $         & 9.0$\pm$3.0 & 5.0$\pm$3.7 & $ $         & 6.8$\pm$0.6 & 19.5$\pm$0.3 \\
   \hline
  \end{tabular}
 \end{small}
 \caption{\label{tab: Flyby 2016 slopes} Selection of spectral slope values in 
    the 535--743 nm range: selection of values from the measurements marked in 
    the right panel of figs.\ref{fig: Flyby 2016 panels}, 
    \ref{fig: Flyby 2016 I/F}, and \ref{fig: Flyby 2016 3 filters spectro}. 
    The spectral slopes of the bright features (BF) are almost one-third of 
    those of the smooth regolith (UR) and just less than one-fourth of those 
    of the somber boulders (SB).
    }
\end{table*}
In a similar manner but at the opposite end of reflectances, the boulders 
exhibiting the red spectral behavior that we discussed above also distinguish 
themselves from their surroundings in these reflectances mappings through their 
lower-than-average reflectance. They present the same behavior independently 
of the phase angle at which they are observed.
\subsection{Spectrophotometric analysis of local features}
\begin{figure*}
 \begin{minipage}[b]{\linewidth}
  \begin{minipage}[b]{\linewidth}
    \begin{center}
    \includegraphics[width=0.97\linewidth]{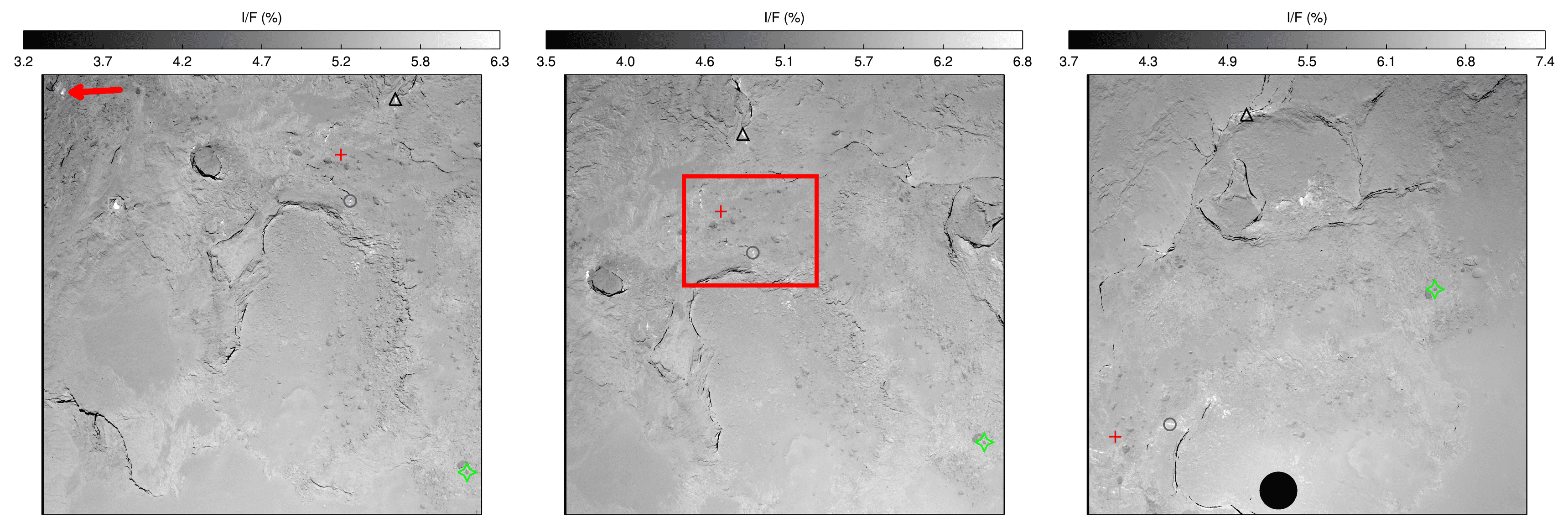}
    \end{center}
    \caption{Panel of the photometrically corrected radiance factor
       images taken with the orange filter. From left to right: Image
       taken at 23h34, 23h46 and 00h01. The average phase angle is
4\dg, 3\dgr, and 1\dg, respectively. See fig. \ref{fig: Flyby 2016 panels}
       for the symbol keys. The red square in the central figure 
       points to the area observed in August 2016 that is shown in fig. 
       \ref{fig: august 2016}.}
    \label{fig: Flyby 2016 I/F}
  \end{minipage}
  \begin{minipage}[b]{\linewidth}
    \begin{center}
    \includegraphics[width=0.97\linewidth]{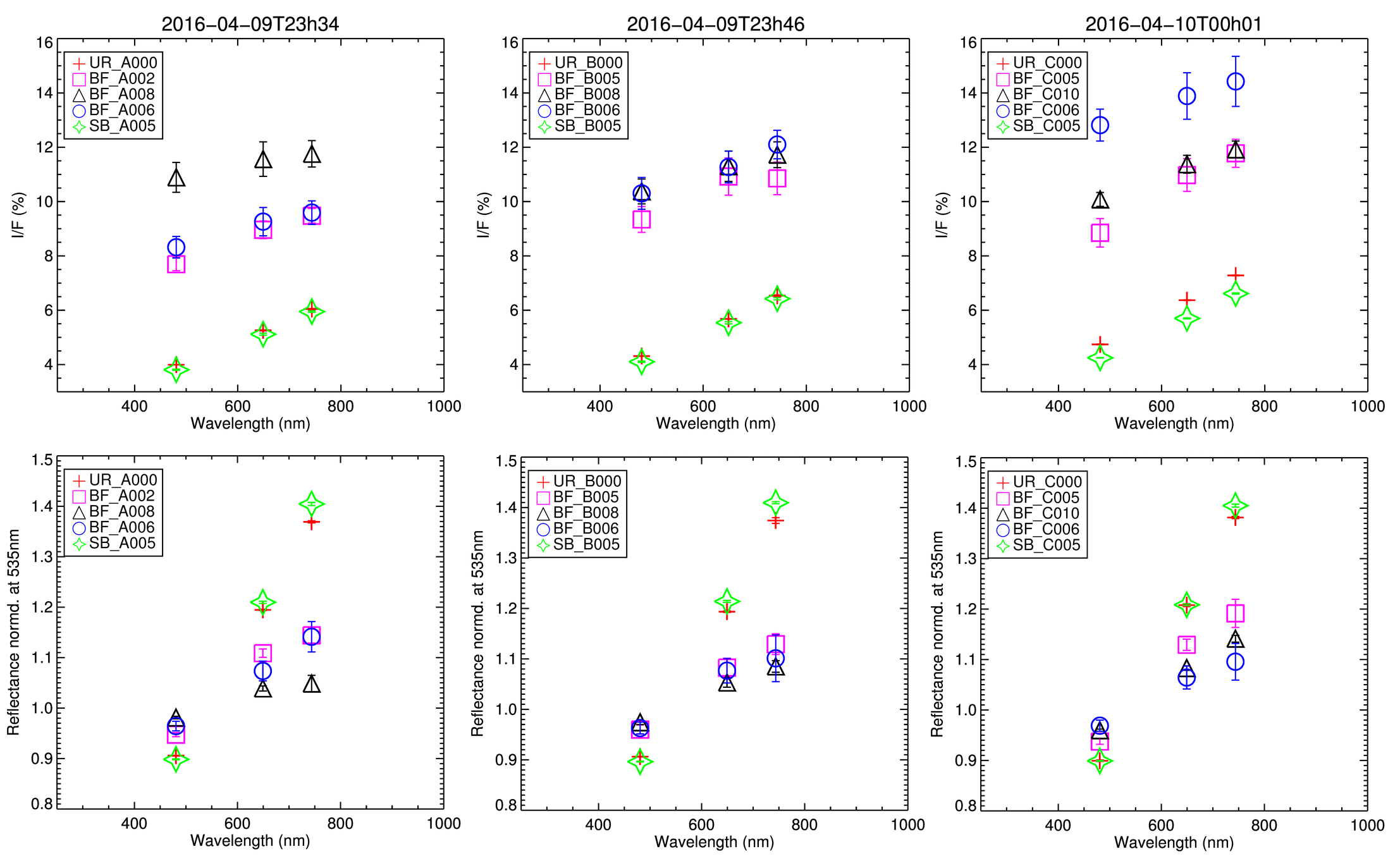}
    \end{center}
    \caption{Panel of the result of the three-filter spectrophotometry measurements. 
       Top: Reflectance (I/F) measurements. The bars indicate the 
       dispersion of the values measured within the box and their propagation for 
       the relative reflectance. Bottom: I/F measurements normalized at 535 nm. 
       The key for the measurements IDs is, again, as follows: smooth, unresolved-%
       looking, regolith (UR), bright feature (BF), and somber boulder (SB). The 
       letter after the underscore indicates the 23:34 image (A), the 
       23:46 image (B), or the 00:01 image (C). Finally, the last three digits 
       identify the feature among its particular type. See fig. \ref{fig: Flyby 2016 I/F} 
       for an illustration of the measurement locations.}
    \label{fig: Flyby 2016 3 filters spectro}
  \end{minipage}
 \end{minipage}
\end{figure*}
As this way, several dozens of surface elements were identified in the images as 
particular features of interest, for instance, bright patches, somber boulders, 
consolidated material, and unresolved regolith. From this collection of elements, 
we selected six features appearing in figs. \ref{fig: Flyby 2016 panels}, 
\ref{fig: Flyby 2016 I/F} and \ref{fig: Flyby 2016 3 filters spectro}.\\
The radiance factor and spectral slope measurements were performed by 
integrating the signal in a 3x3 pixels box (i.e., 1.6 x 1.6 m$^{2}$), unless 
otherwise noted. These measurements were made on both common and image-unique 
features. All images presented here have three measurements in common: a stretch 
of smooth regolith (red cross), a bright spot underneath a overhang (blue 
circle), and a somber boulder (green star). These measurements are identified 
by a tag: UR for very smooth unresolved-looking regolith, BF for bright 
feature, or SB for somber boulder (see fig. 
\ref{fig: Flyby 2016 3 filters spectro} and tables 
\ref{tab: Flyby 2016 stats} and \ref{tab: Flyby 2016 slopes}).\\
We show the results of the three-filter spectrophotometry in fig.
\ref{fig: Flyby 2016 3 filters spectro}.\newline\\
Based on these measurements, we pursued the previously discussed notable 
distinction in behavior between the selected bright features 
and the other regions of interest. While somber boulders and smooth 
unresolved-looking regolith exhibit a similar behavior in both radiance factor
and relative reflectance, all the selected bright features have a radiance 
factor at least twice that of the somber boulders or the unresolved regolith 
(see fig. \ref{fig: Flyby 2016 3 filters spectro} and table 
\ref{tbl: spectral slopes tbl}). Moreover, the bright surfaces also 
display smaller and less red spectral slopes than those of other surfaces and 
differ more widely from the slope of the average terrain. For the investigated 
terrains, for instance, while the spectral slopes of the somber boulders are 
between 1\% and 19\% higher than the slopes of the smooth regolith, 
the spectral slopes of the bright surfaces are between 10\% and 87\% smaller 
than those of the smooth-looking regolith. All corresponding values are 
assembled in table \ref{tab: Flyby 2016 slopes}.\newline\\
In the flybly area of February 2015 (encompassed by the green 
dashed line in the left panel of  fig.\ref{fig: context flyby}), some small bright surfaces 
were observed along the decline that leads toward the center of the Imhotep 
depression. However, these bright surfaces had reflectivities that would never 
be higher than 50\% than the local average, and their spectral slopes were as 
high or greater than those of neighboring smooth terrains. These surfaces were 
interpreted to be partially richer in refractive materials and coated by 
deposits of organics \citep{Feller_2016}.\\
Given the wider range of variations of the reflectance and the flatter 
spectra exhibited by the bright surfaces observed here, we interpret these 
surfaces to be of a similar nature to some of the surfaces investigated in 
 \citet{Pommerol_2015, Filacchione_2016a, Barucci_2016} and \citet{Oklay_2016}, which 
were observed with the OSIRIS and VIRTIS instruments and were shown to be 
slightly richer in water-ice content, but only by a few percent. 
In this understanding, the more water-ice a surface element of the nucleus 
contains, the flatter its spectrum. A comprehensive study of such 
surfaces and of tentative modeling of their composition has been presented 
in \cite{Raponi_2016}.\\
Following the conclusions of these papers, we consider the surfaces to be appropriate 
candidates as locations enriched in water-ice material at the time of these 
observations. These locations should be further investigated and compared with 
possible VIRTIS observations.
\begin{figure*}
 \begin{center}
  \includegraphics[width=0.70\linewidth]{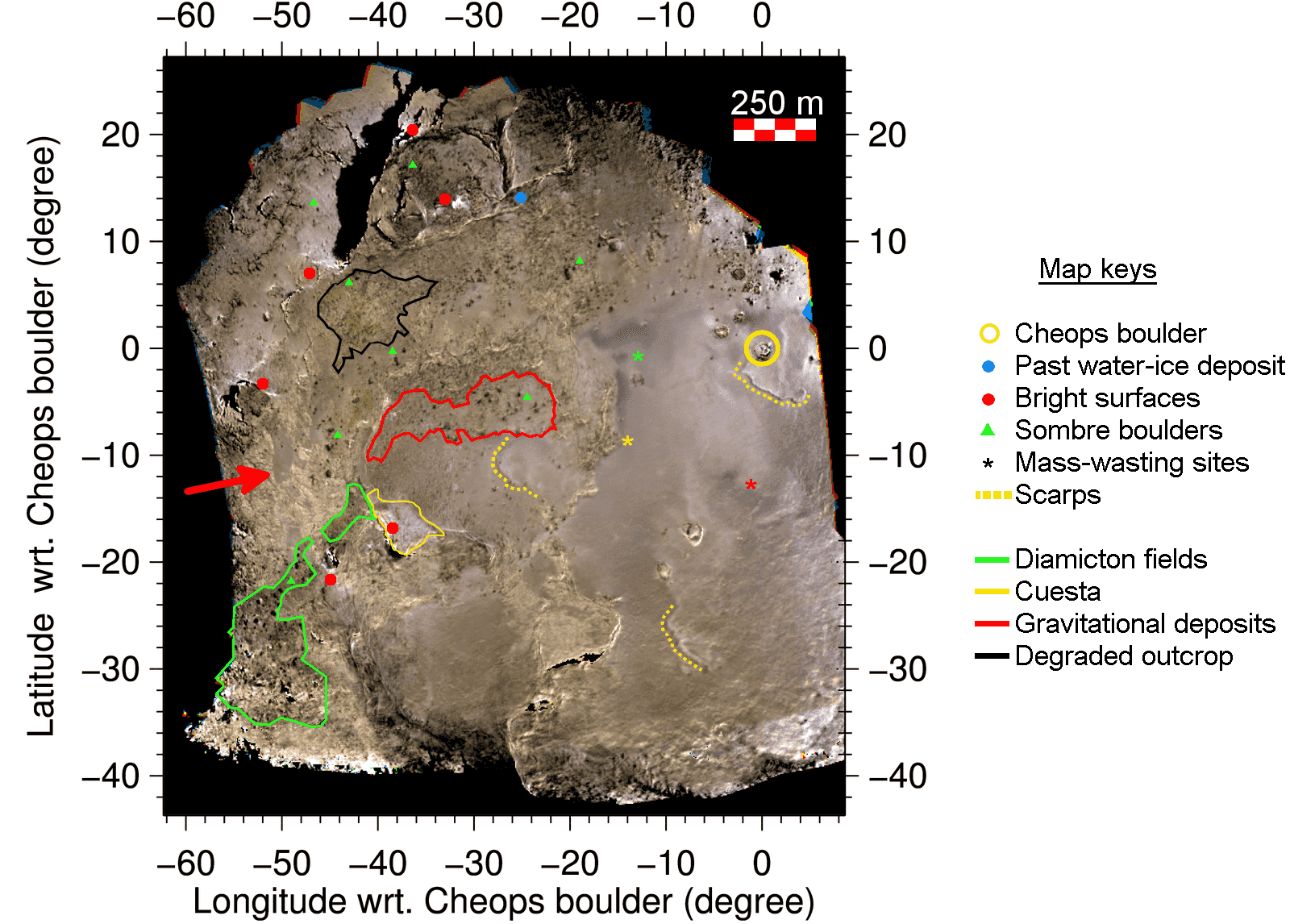}
 \end{center}
 \caption{Georeferenced RGB of the April 2016 flyby region, assembled using 
    photometrically corrected NF84, NF82, and NF88 images, and centered around 
    the Cheops boulder. This mapping has a 0.1\dg x0.1\dgr resolution in 
    longitude and latitude. The region shown correspond to areas visible in at 
    least three of the considered observations. The locations where resurfacing 
    processes were observed through the evidence of mass-wastings and moving 
    scarps that were discussed in e.g. \protect\citet{Groussin_2015} and 
    \protect\citet{Deshapriya_2018} are indicated. For clarity, 
    only the largest bright spots are indicated, and the given positions of 
    somber boulders point to their presence in all type of terrains, with the 
    exception of the central depression of Imhotep.}
 \label{fig: RGB GEOTIFF} 
\end{figure*}
\subsection{Georeferenced RGB mapping}
As the previously presented RGBs only allowed us to consider part of the 
flyby region as it was around the moment of closest approach, we have 
projected each photometrically corrected image onto a grid and assembled the 
results in a georeferenced RGB of the area (see fig. 
\ref{fig: RGB GEOTIFF}).\\
To produce this mapping, we have generated the georeference of each  
image to the SPG SHAP7 v1.0 model using ray-tracing and the reconstructed 
trajectory of the Rosetta spacecraft with respect to landmarks on the nucleus, 
computed during the generation of the SPC shape model.\\
We then projected the images onto a grid with a 0.1\dg x 0.1\dgr resolution in 
longitude and latitude, and averaged the reflectance by the number of times a 
grid element appeared in the observations. This resolution in longitude and 
latitude corresponds to a resolution of (2.7$\pm$0.3)x(2.7$\pm$0.3) m$^2$.%
\newline\\
In this figure, the areas encircled by continuous lines correspond to some of 
the morphological units defined in fig. \ref{fig: geomorpho}, the red circles 
and green triangles identify some of the bright surfaces and 
somber boulders, respectively. Scarps are noted here by the dotted 
yellow line, while a well-defined fine-dust deposit surrounded by consolidated 
material is indicated by the red arrow. Furthermore, the colored stars 
indicate areas pointed out by \citet{Groussin_2015} as origins of resurfacing 
processes in the weeks before the comet passage at perihelion, while the 
blue circle indicates the position of pre-perihelion bright spots first 
discussed in \citet{Pommerol_2015}, where water-ice was
detected \citep{Filacchione_2016}.\\
The variations in color and hue displayed in fig. \ref{fig: RGB GEOTIFF} 
are a compelling indication of the multiplicity of terrain types. We recall 
that in a figure like this, a relative difference in brightness between 
two surface elements is indicative of a variation in geometric albedo, while 
a relative difference in color is indicative of a distinction of spectral 
behavior. We selected here and designated a few elements discussed 
previously, as well as in previous studies of this area.\newline\\
While not all bright spots are apparent with this resolution, some of 
the bright surfaces (red circles in fig. \ref{fig: RGB GEOTIFF} and in the 
previous figures) are still easily distinguishable. Those pointed here 
are the largest and located at the bottom of underhangs, except for two: 
one located around (-32\dg, +14\dg), found at a base of a niche of the Ash 
region, and another located close to (-46\dg, -22\dg) at the feet of a 
megaclast.\\
We note here that these two surface elements are both part of the source 
regions of two activity events that occurred around perihelion in August 2015.
Both events are discussed in \citet{Fornasier_2018}, and are numbers 42 and 
133 in table A1.\newline\\
Additionally, we also refer to \citet{Oklay_2016a} for the investigation of 
the (-32\dg, +14\dg) bright feature (denoted there ROI7) in pre-perihelion 
images (NAC filter sequence acquired at 2014-09-25 06:46:25 UTC). This bright 
feature was then notably compared to the area marked by a blue circle (-26\dg, 
+14\dg), which was also found to harbor particularly bright boulders and 
surface elements whose spectral slope was lower than their surroundings 
(\citealp{Pommerol_2015} and \citealp{Oklay_2016a}). This particular area, 
observed by the VIRTIS instrument in October 2014 and labeled BAP-1 in 
\citet{Filacchione_2016}, was found to have spectral properties that are best 
fit with a water-ice content that varies between 1.2\% and 3.5\%, depending on 
the mixing scenario the authors considered (areal or intimate mixing), and are 
therefore one of the very first nucleus regions where emerging water-ice-rich 
material was ascertained.\newline\\
The two bright surfaces, located around (-44\dg, +8\dg) and (-38\dg, +20\dg), 
are both particularly evident in the spectral slope mappings of the bottom left 
panel in fig. \ref{fig: Flyby 2016 panels}, and BF-010 and BF-011 have 
geometric albedos at 649 nm of 11.3\% and 8.9\%, respectively (i.e., $\sim$ 
66\% and 31\%, respectively, higher than the nucleus average of 6.8\%) and 
spectral slopes of 6.8\% and 9.0\% (i.e., $\sim$ 62\% and $\sim$51\%, 
respectively, lower than the region average of 17.98 \%/100 nm at $\alpha$ = 
0\dg, see fig. \ref{fig: phase reddening}). These two surfaces are found at 
the bottom of large declivities, and the nature of the topography between the 
morphological regions Ash and Aten is only well rendered in fig. 
\ref{fig: RGB GEOTIFF} by the actual absence (areas in black) of observations 
of these cliffs during the flyby manoeuvre.\newline\\
For similar reasons, the bright surface located around -55\dg of longitude and 
-2\dg of latitude, identified in this study as BF-08, also stands out because 
during the 23h34 and 23h46 observations, its reflectance was at higher than 
11\%, while its spectral slope ranged between 0.8 and 5.1\%/10 nm (i.e., 
between $\sim$72\% and $\sim$96\% lower than the region average).\\
Following the previous remarks, it is therefore most likely that these 
bright surfaces harbor water-ice-rich material.\newline\\
Similarly, in this mapping, the largest somber boulders are also still 
distinguishable in different parts of the region (see the green triangles in 
fig. \ref{fig: RGB GEOTIFF}). They are found to be on most types of 
morphological units: diamictons (green polygons), fine-material deposits 
covered with boulder fields, fine-material  deposits (e.g., the lone boulder in 
Aten in between the two bright surfaces), as well as among the gravitational 
accumulation deposits (red polygon) and the degraded outcrops (black polygon). 
We note, however, that they are absent from the fine-material deposits that cover 
the smooth central region of the Imhotep depression, as well as from two of the 
taluses of the area.\\
We note here that such boulders with a lower-than-average reflectance and a 
red spectral behavior have also been observed in the area of the February 2015 
flyby \citep{Feller_2016}.\newline\\
Other areas of this flyby region present a contrast as clear as that of somber 
boulders and their neighboring terrains, for instance, the area indicated by 
the red arrow in fig. \ref{fig: RGB GEOTIFF}, where previous fine-material 
deposits cover the top of a series of terraces whose fronts are formed by the 
outcropping consolidated material. Moreover, scarps within fine-material 
deposits are clearly visible. They are denoted by the yellow arrows in the 
corresponding figure. Such features were evident in pre-perihelion images, for 
instance, the red star marks the linear feature shown in \cite{Thomas_2015a}. 
During the approach to perihelion, however, some new scarps appeared or 
progressed across the surface, exposing resurfacing processes of the smooth 
central area of Imhotep \citep{Groussin_2015}, with evidence of mass-wasting 
processes around the location marked by the yellow, green, and red stars as 
well. These resurfacing processes are also discussed in \cite{Deshapriya_2018}. 
We also note here that the area around the red star was the source from 
another activity event observed in NAC observations acquired on 12 August 2015 
(see entry 151 in table A1 of \citealp{Fornasier_2018}).\\
While the contrast between the top and bottom of these scarps is well marked 
in this figure and the outline of the scarp close to (-20\dg, -10\dg) 
is even evident at a low phase angle (see the right panel of fig. 
\ref{fig: Flyby 2016 I/F}), the differences in the corresponding spectral 
slope mapping are notably small (see fig. \ref{fig: Flyby 2016 central slp}) as 
it presents itself as a variation of less than 2 \%/100 nm around the local 
average spectral slope value.\newline\\
The flyby area is distinguished by the diversity of morphological units, as 
well as by the variety of colors of these different terrains and of particular 
features such as bright spots and somber boulders. Furthermore, this 
particular region of the comet can be deemed of a particular interest as it 
encompasses numerous surfaces elements where cometary activity and surface 
evolution have been observed as the comet approached, passed through 
perihelion, and moved away from it. The source locations of jets and outbursts 
observed around perihelion are shown in fig. 1 of Fornasier et al., submitted.
For instance, the area where niches hosts two large bright surfaces located 
around (-32\dg, 14\dg) was observed on 1 August 2015 to be the source of 
one particular large outburst. Such bright surfaces were not evident in 
OSIRIS/NAC images acquired in September 2014 (\citealp{Auger_2015} and 
\citealp{Oklay_2016}), but further indicate compositional heterogeneities on 
the nucleus immediately below the dust mantle, that are revealed as the 
insolation is sufficient to pierce through this insulating layer and trigger 
a violent activity event, such as has been observed for cliff collapse 
(\citealp{Pajola_2016} and \citealp{Fornasier_2016}).
\FloatBarrier
\subsection{NUV, Vis, and NIR spectrophotometry}
\label{sec:section_4_4}
\begin{figure*}
 \begin{minipage}[t]{0.47\linewidth}
  \begin{center}
   \includegraphics[width=\linewidth]{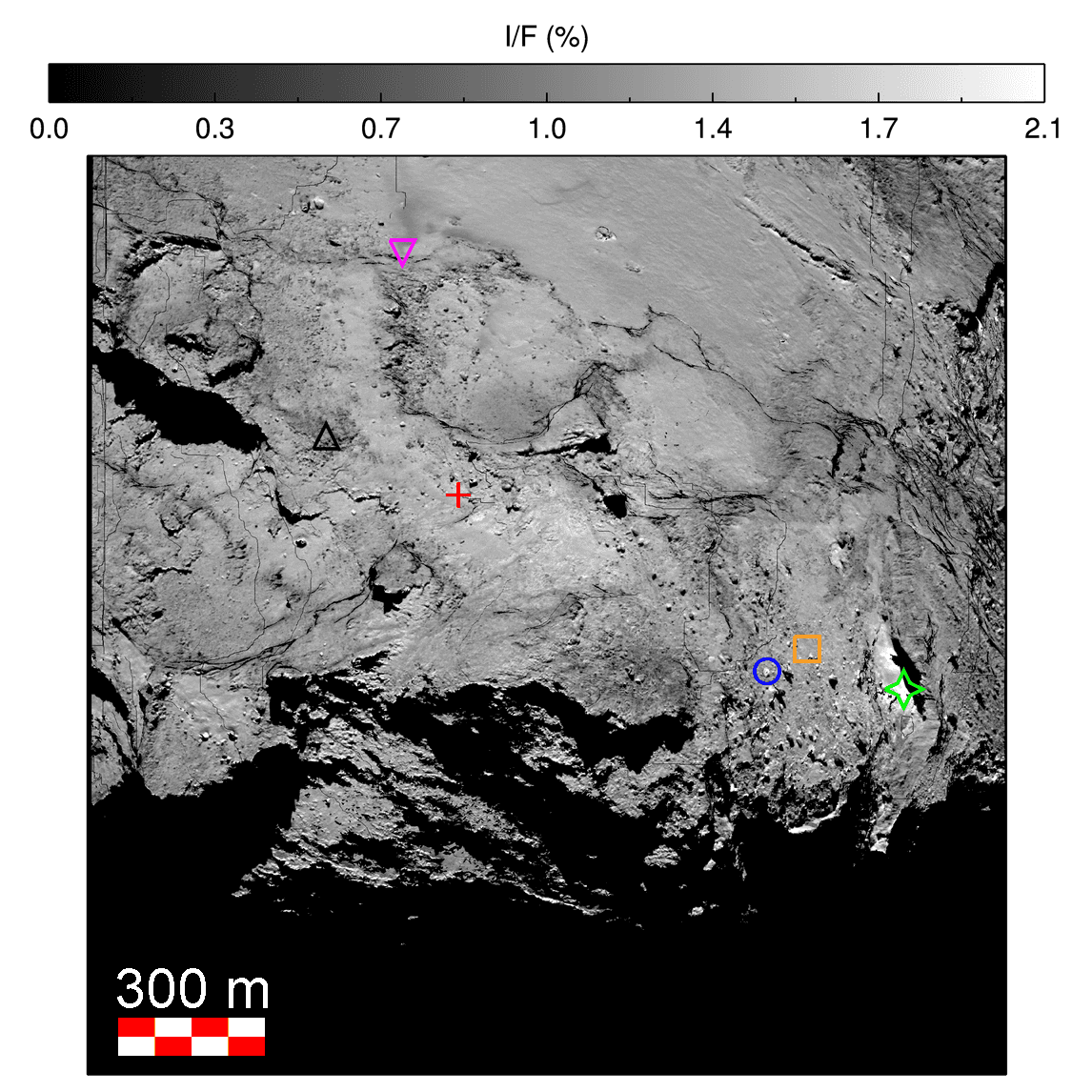}
  \end{center}
 \end{minipage}
 \begin{minipage}[t]{0.47\linewidth}
  \begin{center}
   \includegraphics[width=\linewidth]{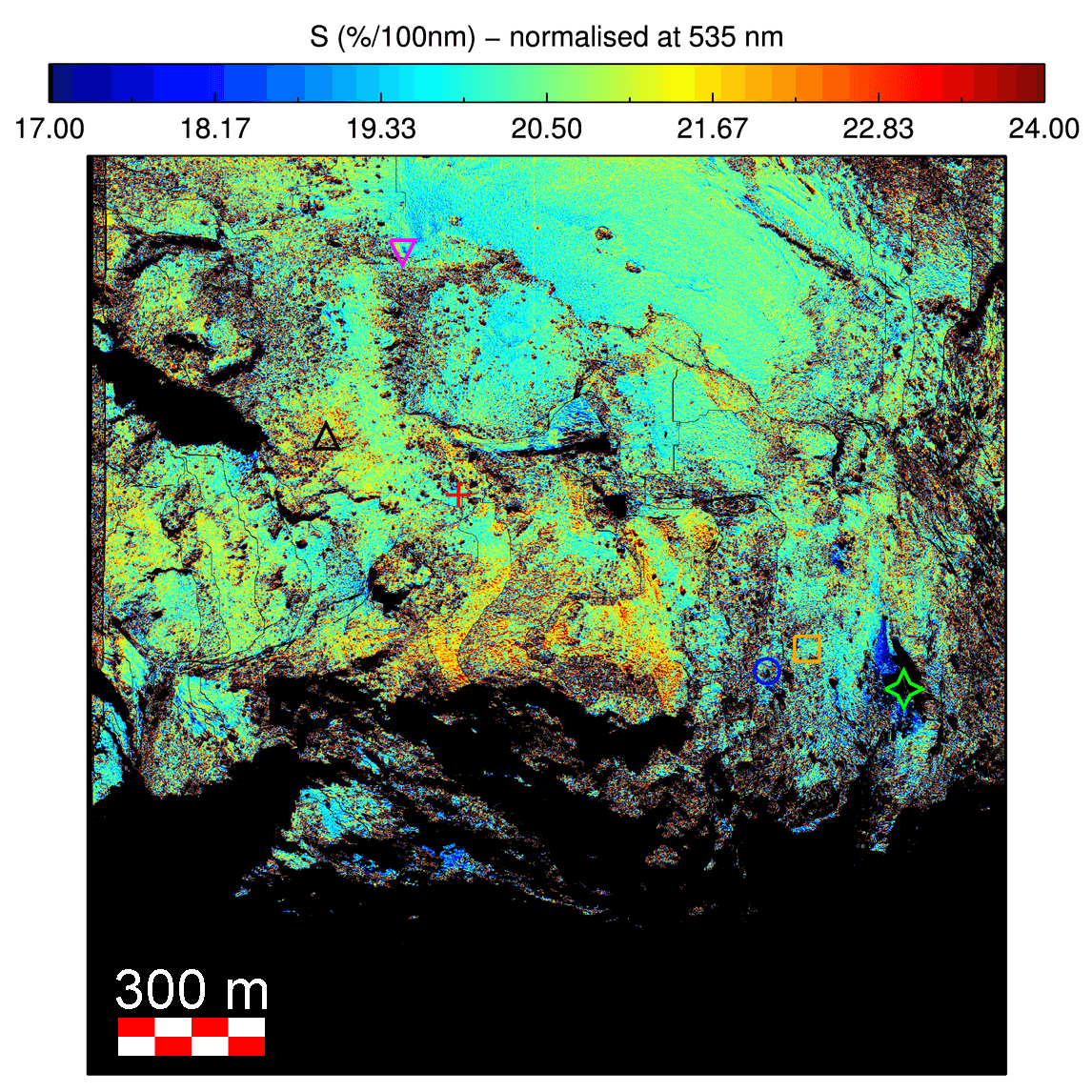}
  \end{center}
 \end{minipage}
 \caption{2016-04-10T11h48 sequence ($\alpha$ $\sim$ 51\dg): Maps of the 
    photometrically corrected radiance factor (left) and of the spectral slope 
    in the 535--882 nm range (right). The symbols refer to the measurements 
    made and presented in fig. \ref{fig: Flyby 2016 11f spectro}. Under 
    these illumination conditions  ($\alpha$ $\sim$ 51\dg), the shadows 
    are cast on the niches and areas where bright material was previously 
    observed (see fig. \ref{fig: Flyby 2016 panels}).}
 \label{fig: flyby 2016 11f ima}
\end{figure*}

\begin{figure*}
 \begin{center}
  \includegraphics[width=0.9\linewidth]{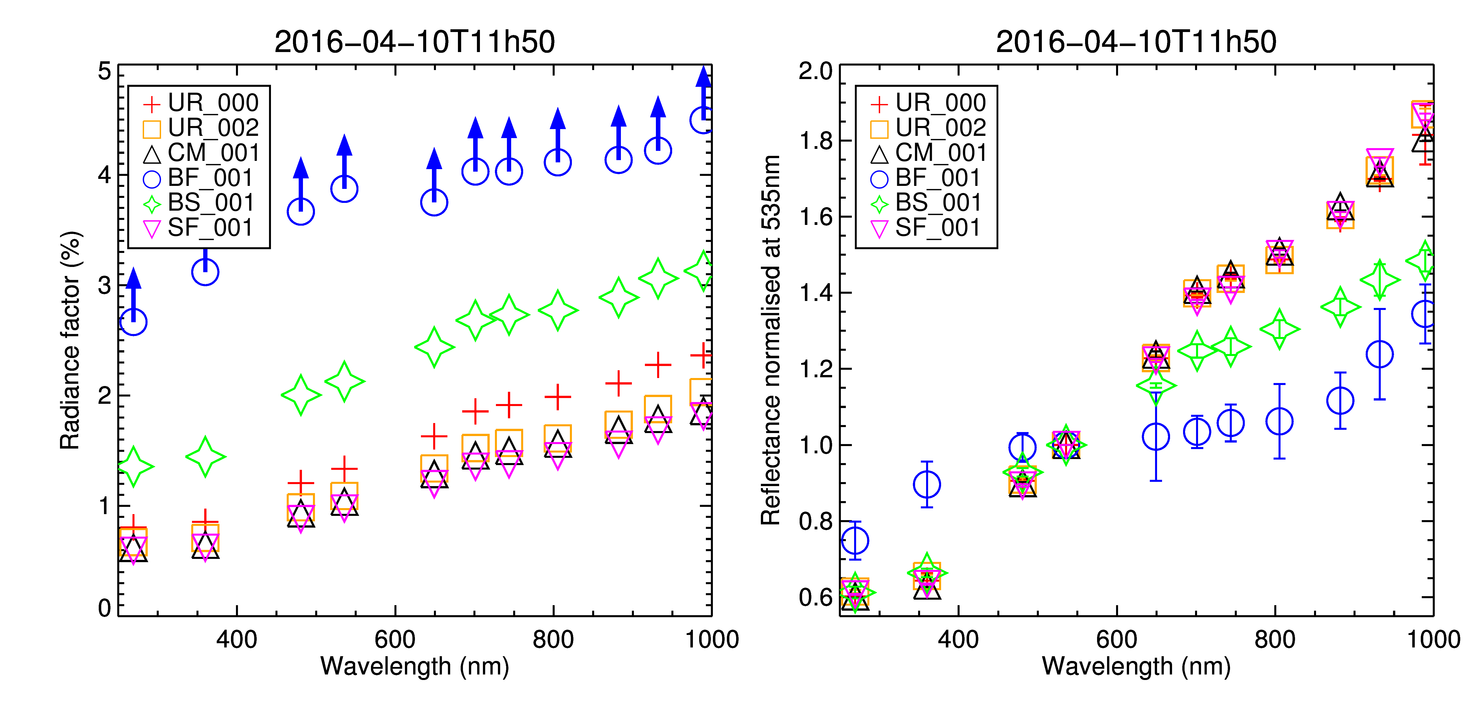}
 \end{center}
 \caption{Spectrophotometry measurements for the 2016-04-10T11h48 sequence. The 
    images were photometrically corrected using the Lommel-Seeliger disk law. 
    The symbols correspond to the measurements made at the locations indicated 
    in fig. \ref{fig: flyby 2016 11f ima}a. We note here that the measures for 
    BF-01 were close to the detector saturation levels. These values therefore 
    correspond to lower limits.}
 \label{fig: Flyby 2016 11f spectro}
\end{figure*}

\begin{figure}
   \begin{center}%0.84:printer %0.60:referee
   \includegraphics[width=0.84\linewidth]{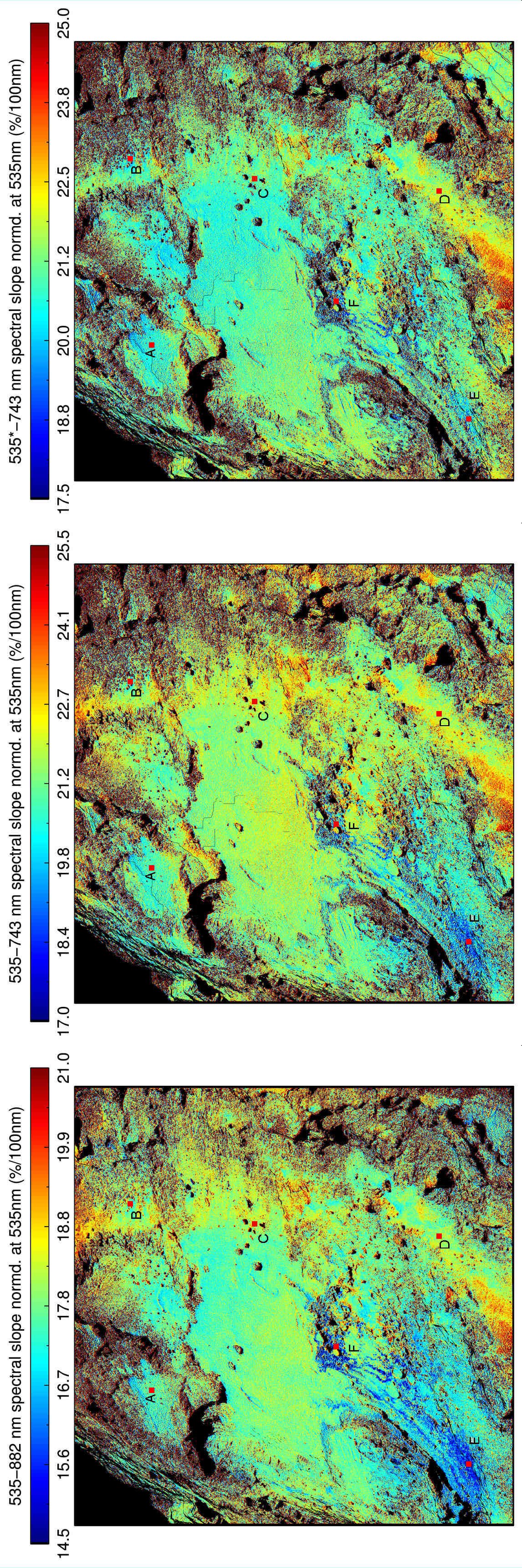}
   \end{center}
 \caption{From bottom to top: Spectral slopes computed in the 535-882 nm
    range, in the 535-743 nm range, or in the 535-743 nm range with an
    interpolation at 535 nm. Each spectral slope mapping has its own color bar,
    adjusted within $\pm$7$\sigma$ around the median value.}
 \label{fig: spc slp 12h13}
\end{figure}

We present here the results acquired from the two sequences of observations
taken before and after the moment of closest approach using the 11 NAC filters
between 269 and 989 nm.\\
These two sequences were found to be best coregistered when segmentation and 
optical flow algorithms were combined. Considering the difficulty of correcting 
the parallax effect for the full image, we chose to integrate the signal in 5x5 
pixels boxes, which corresponds to a surface element of 24.5 m$^{2}$. We used 
both sequences to investigate some particular surface features as well as to 
constrain the differences of spectral slopes described previously.\newline\\
We present in fig. \ref{fig: flyby 2016 11f ima} the reflectance mappings and 
mappings of the spectral slope computed in the 535-882 nm range for the 
2016-04-10T11h50 observation sequence. As in previous observations, we note 
the same apparent anticorrelation between the brightness (in radiance factor) 
of an area, its morphological type, and the steepness of its spectral slope.\\
In these figures, areas that were previously identified as fine-material 
deposits (such as terraces and surfaces close to the bottom of underhangs) 
appear slightly brighter and present a smaller spectral slope. Similarly, the 
consolidated material structure encircled in red in the bottom panel of fig. 
\ref{fig: Flyby 2016 panels}, although it has a reflectance that is almost 
average, still exhibits, at $\sim$ 51\dg of phase angle, a spectral slope 
slightly lower than 17 \%/100 nm. This observation supports the previously 
stated conclusion that this feature likely is a large-scale (of about 100 m) 
compositional heterogeneity.\\
Meanwhile, the structures that are identified as eroded consolidated material,
diamictons, or boulder fields here also present a reflectance that is slightly 
lower than average and a higher spectral slope.\\
In these figures, one feature that is not visible in other NAC images is the 
large bright surface close to the underhang with a low spectral slope. This 
area is also distinguished through its particularly smooth appearance. This 
feature was one of the bright spots whose photometric properties were 
investigated in \cite{Hasselmann_2017} and were found to best match the 
remnants of a sublimated intimate mixture of water-ice, carbon black, and 
tholins studied in \citet{Jost_2017a}. We further investigate its 
spectrophotometric properties below.\newline\\
In this sequence, we investigated the bright spot on the boulder that were 
previously identified (BF-01, blue circle), a neighboring unresolved-appearing 
terrain (orange square, UR-02), a bright smooth surface underneath an overhang
(green star, BS-01), the previously investigated smooth-looking regolith (red 
cross, UR-00), a smooth area in a degraded outcrop of consolidated material 
(black upward triangle), and a somber feature in fine-material deposits on 
Imhotep (magenta downward triangle, SF-01). We report the results of the 
spectrophotometric analysis in fig. \ref{fig: Flyby 2016 11f spectro}.%
\newline\\
We report here the values of the spectral slopes in the 535--743 nm range for
the measurements that were made without correcting for the phase-reddening
phenomenon (see following section). These measurements are (in
\%/100 nm) 20.4$\pm$0.4 for UR-00, 20.0$\pm$0.9 for UR-02, 20.9$\pm$0.4
for CM-01, 3$\pm$6 for BF-01, 12$\pm$2 for BS-01, and 19.2$\pm$0.7 for
SF-01.\newline\\
Figure \ref{fig: Flyby 2016 11f spectro} shows that the smooth-looking 
terrain and the consolidated material exhibit a similar red spectral behavior, 
as has been observed previously in the high-resolution images of the February 
2015 flyby. This spectrum for the average smooth-looking regolith terrain 
(UR-00) is typical of the average dark terrain of the comet nucleus (see, e.g., 
fig. 2 in \citealp{Fornasier_2016}). However, contrary to what was observed in 
February 2015, the bright surfaces of this area (BF-01 or BS-01) have spectra 
sensibly different from smooth surfaces or consolidated material. Moreover, the 
selected bright areas are at least twice as bright (in terms of radiance 
factor) as the average smooth-looking regolith terrain.\\
We indicate here that in all of these observations, the reflectances of the
bright feature at the top  of a boulder (called BF-01 in this study and ID-44 in
\cite{Deshapriya_2018}) were again similar to but did not reach the saturation 
levels of the NAC. This is denoted by the blue arrows in the left panel of fig.
\ref{fig: Flyby 2016 11f spectro}.\\
We observe, as noted previously with the three-filters observations, that the
smooth unresolved-looking regoliths, the surfaces among the eroded outcrop, and
the somber feature exhibit similar characteristics: the reflectance is considerably 
lower than those of the bright features, and they share a very similar spectral 
behavior. On the other hand, the two identified bright features have very low 
spectral slopes at $\sim$51\dg  \ phase angle and noticeably fainter normalized 
reflectances. We interpret these surfaces to be fractionally enriched in water-%
ice material (\citealp{Filacchione_2016a, Barucci_2016, Deshapriya_2016, 
Fornasier_2016} and \citealp{Oklay_2016}) , and the observed differences between 
these two surfaces might reflect a further difference in composition, as 
considered in \citet{Deshapriya_2018}.\newline\\
We also considered these two sequences with 11 filters in order to constrain
the differences in spectral slopes using different computing methods.\\
In previous studies, spectral slopes have been computed in the 535-882nm range
when the corresponding OSIRIS images were available. Spectral slopes have
otherwise been computed in the 535-743 nm range, using either the F23 filter 
image, or an interpolation using the F24 and F22 filter images. We present in 
fig. \ref{fig: spc slp 12h13} the corresponding mapings.\newline\\
We have investigated the global variations of the spectral mappings assembled
from the 2016-04-09T12h13 and 2016-04-10T11h50 observations, as well as some 
local variations between the assembled 2016-04-09T12h13 spectral slope mappings.\\
For each observation, we applied a Gaussian fit to the different spectral slope
mappings to determine the center of the distribution and its full width at half-%
maximum (FWHM). The results are assembled in table \ref{tab: spc global}.\\
\begin{table}
 \centering
 \begin{tabular}{|c|c|c|}
  \hline
  \hline
  Observation & $\Delta\lambda$ (nm) & S (\%/100 nm)\\
  \hline
  9$^{th}$/04 - 12h13   & 535 - 882  & 17.8$\pm$2.2 \\
  $\alpha$ $\sim$ 62\dg & 535 - 743  & 21.6$\pm$2.9 \\
  $ $                   & 535* - 743 & 21.2$\pm$2.6 \\
  \hline
  10$^{th}$/04 - 11h50  & 535 - 882  & 17.6$\pm$2.3 \\
  $\alpha$ $\sim$ 51\dg & 535 - 743  & 21.0$\pm$3.0 \\
  $ $                   & 535* - 743 & 20.4$\pm$2.6 \\
  \hline
 \end{tabular}
 \caption{\label{tab: spc global} Differences in spectral slopes: spectral 
   slope values (S, center, and FWHM) of the Gaussian fitting for each 
   spectral slope mapping considering their wavelength range ($\Delta\lambda$). 
   The asterisk  denotes that the corresponding 535 mappings were interpolated 
   from the associated 480 nm and 649 nm filter images.
   }
\end{table}
We note that based on these results for the 12h13 observations, the spectral
mapping in the 535-882 nm range is  $\sim$21\% and $\sim$19\% lower than the 
535-743 nm and the 535* -743 nm spectral mappings. Similarly, for the 11h50 
observations, the spectral slopes mapping in 535-882 nm range is $\sim$19.5\% and 
$\sim$16.5\% lower than the other two spectral mappings.\newline\\
\begin{table}

 \centering
 \begin{tabular}{|c|c|c|c|}
  \hline
  \hline
  Location & \multicolumn{3}{c|}{Spectral slope range (nm)} \\
  $ $                & 535 - 882      & 535 - 743       & 535* - 743 \\
  \hline
  ROI - A            & 17.7 $\pm$ 0.7 & 21.2 $\pm$ 0.5 & 20.8 $\pm$ 0.5 \\
  ROI - B            & 17.2 $\pm$ 0.2 & 20.3 $\pm$ 0.4 & 20.0 $\pm$ 0.4 \\
  ROI - C            & 18.5 $\pm$ 0.5 & 22.3 $\pm$ 0.5 & 21.6 $\pm$ 0.7 \\
  ROI - D            & 18.1 $\pm$ 0.2 & 21.6 $\pm$ 0.2 & 21.7 $\pm$ 0.3 \\
  ROI - E            & 14.7 $\pm$ 0.2 & 19.5 $\pm$ 0.7 & 20.8 $\pm$ 0.5 \\
  ROI - F            & 15.2 $\pm$ 0.3 & 19.1 $\pm$ 0.4 & 18.9 $\pm$ 0.4 \\
  \hline
 \end{tabular}
 \caption{\label{tab: spc local} Comparison of the spectral slope ranges:
    Different spectral slope values for the different regions of
    interests indicated in fig. \ref{fig: spc slp 12h13}.}
\end{table}
We further list in Table \ref{tab: spc local} local measurements of the
spectral slopes around the Imhotep region, including areas outside the
Imhotep-Khepry transition. Following the previous remark, we find once again
that the 535-882 nm spectral slope measurements are on average 20$\pm$5\%
lower than the 535-743 nm spectral slopes, or 18$\pm$9\% lower than those
computed between 535nm (interpolated) and 743nm.\\
In both cases, the extreme values come from the measurements of the strikingly
blue regions (i.e., those with a lower slope than average) in fig.
\ref{fig: spc slp 12h13}. These 535-882nm measurements differ by 32\% and 26\% 
from the corresponding 535-743nm values and by 41\% and 24\% from the 
corresponding 535*-743nm values.\\
These areas are part of a region of the nucleus where water-ice-rich material 
and outbursts have been observed (\citealp{Knollenberg_2016, Deshapriya_2018, 
Oklay_2016} and \citealp{Agarwal_2017}. They also encompass the feature 
described as “blues veins” in \citet{Hasselmann_2017}, whose photometric 
properties were best matched by those of the remnants of an intimate mixture of 
water-ice, carbon black, and tholins after sublimation. The noted spectral 
difference is therefore very likely to derive from a compositional difference. 
As there appears to be an overall shift down of the 535-882 nm spectral slopes 
in the bottom left corner of the mapping with respect to the others, however, we 
note here that this local difference in values might be due to a new flat-field 
correction for the NAC F24 filter (centered at 535nm) that was implemented 
in-flight in late 2015. While this artifact might slightly overestimate the 
noted particularity of this area (by $\sim$9\%  compared to the other two 
measurements), we still correctly observe that this area is different 
from the other neighboring areas (see the other mappings in fig. 
\ref{fig: spc slp 12h13}).
\subsection{Phase reddening}
\begin{figure}
 \begin{center}
  \includegraphics[width=\linewidth]{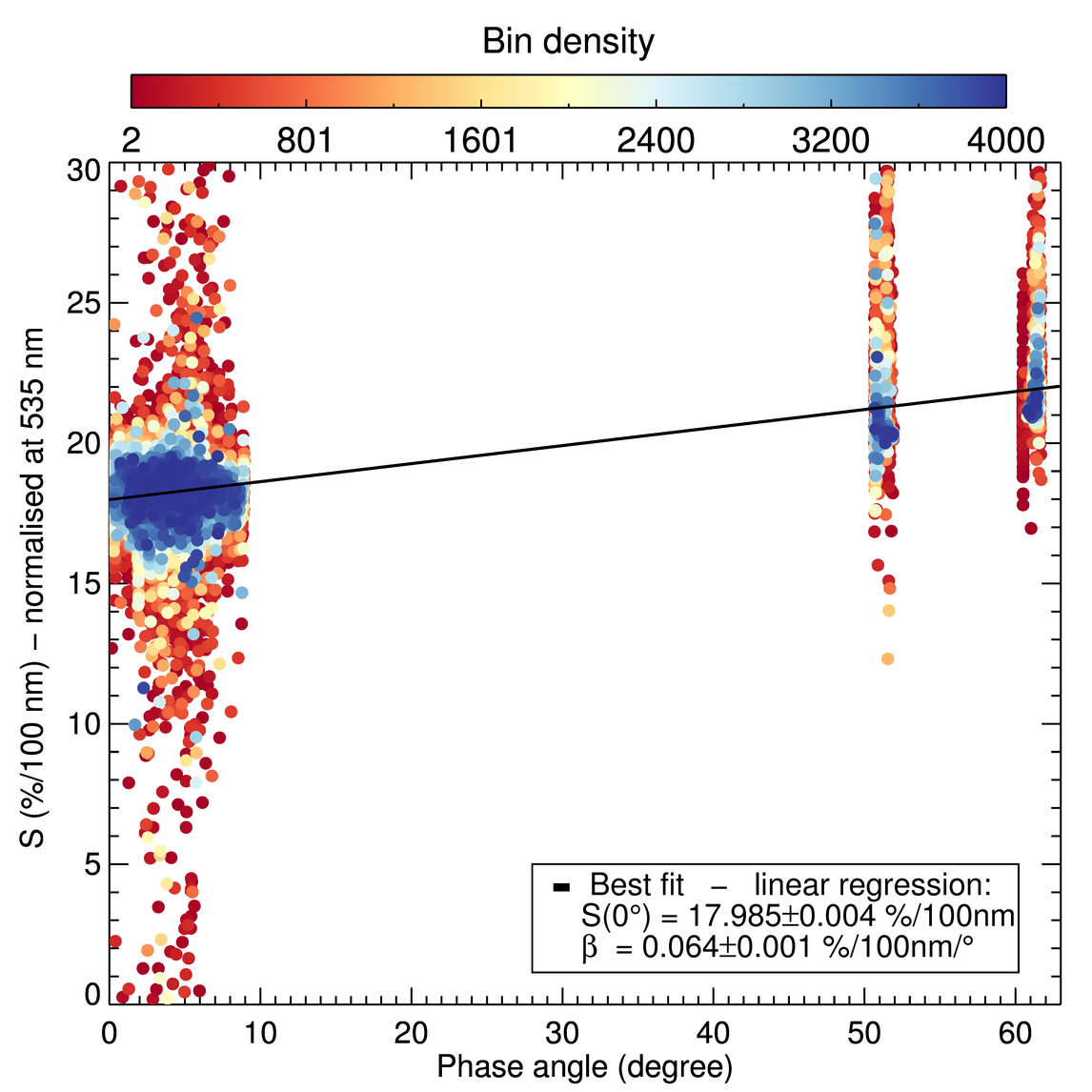}
 \end{center}
 \caption{Phase reddening observed above the Khepry-Imhotep region during the 
    flyby. Using the density of each bin as weights, the linear best-fit solution
    gives a spectral slope at 0\dgr and a phase reddening slope similar to that 
    measured in \citet{Feller_2016}.}
 \label{fig: phase reddening}
\end{figure}
Phase reddening is the increase in spectral slope with the phase angle, 
first observed on the lunar surface \citep{Gehrels_1964}. As discussed at 
length in \citet{Jost_2017a}, this phenonemon has also been observed in 
laboratory settings \citep{Gradie_1980} as well as on asteroids 
(\citealp{Clark_2002} and \citealp{Sanchez_2012}). The ROSETTA mission 
parameters have allowed us soon after rendezvous with the comet to provide the 
first definite phase-reddening measurement on a comet \cite{Fornasier_2015}. 
As in our previous studies (\citealp{Fornasier_2016} \citealp{Feller_2016}), 
we have sought here to constrain the measure of the phase reddening in the 
Imhotep-Khepry transition as the comet was at 2.7 au away from the Sun.\\
To this effect, we computed the spectral slope mappings for each sequence of 
images listed in table \ref{tab: Flyby 2016 NAC images} and then ran a linear 
regression.\newline\\%
The density plot of the spectral slopes computed for this dataset and their
linear best-fit solution are shown in fig. \ref{fig: phase reddening}.\\
Around the morphological transition between Khepry and Imhotep, in the 
535-743 nm range, and between 0.1\dgr and 62\dgr of phase angle, these
spectral slopes are best fit by a slope of 0.064$\pm$0.001 \%/100 nm/\dgr
and a spectral slope at $\alpha$=0\dgr of 17.985$\pm$0.004 \%/100 nm.\newline\\
We note that these results are quite similar to those obtained for the part of 
the Ash-Imhotep transition flyby in February 2015, which were 0.065$\pm$0.001 
\%/100 nm/\dgr and 17.9$\pm$0.1 \%/100 nm, as the comet was 2.3 au inbound to 
perihelion. We recall that the flyby discussed here took place while the 
nucleus was 2.7 au outbound.\\
While the values of these 535-743 nm spectral slopes are $\sim$ 17$\pm$5 \% 
higher than their 535-882 nm counterparts, the phase-reddening slopes are 
nevertheless consistent with those measured in the 535-882 nm range over one 
rotational period of the nucleus, between August 2014 and February 2016, as 
listed in \citet{Fornasier_2016}.\\
As shown in that study, the phase-reddening slope of the nucleus abated from 
0.104 to 0.041 \%/100 nm/\dg between August 2014 and August 2015 (inbound to 
perihelion), before a further increase between August 2015 and February 2016
(outbound from perihelion) at least up to the initial phase-reddening slope 
of 0.104\%/100 nm/\dg. Thermal modeling by \cite{Keller_2015} and analysis of 
detailed nucleus observations by \cite{El-Maarry_2016} have lead to the 
interpretation of this variation as a likely consequence of the global 
thinning of the dust mantle across the nucleus that is observed around perihelion (on 
an order of magnitude of 1 m during this passage), and the change in surface 
properties, such as dust roughness and the composition of the uppermost layer.%
\newline\\
This area, which has been the source region of several jets during that passage 
of the comet through the inner solar system, is also the area where most of the 
fine-material deposits of the Imhotep central area went through a resurfacing 
process in the weeks before perihelion. Likewise, this is the area 
where evidence of diurnal and seasonal water cycles were observed 
\citep{Fornasier_2016}, and where more surfaces likely containing water-ice 
material are exposed on the surface by April 2016 than in August or September 
2014. These surfaces were then, at 2.7 au, likely to survive until the approach 
to the next perihelion passage.\\
We therefore interpret the increase in phase-reddening slope outbound from 
perihelion to reflect the complex nature of the changes that occurred and were 
occurring on this part of the nucleus surface and the transition in terms of 
cometary activity as its heliocentric distance grew.\newline\\
%
%Section discussion
\section{Discussion}
\begin{figure*}
 \begin{center}
  \includegraphics[width=0.75\linewidth]{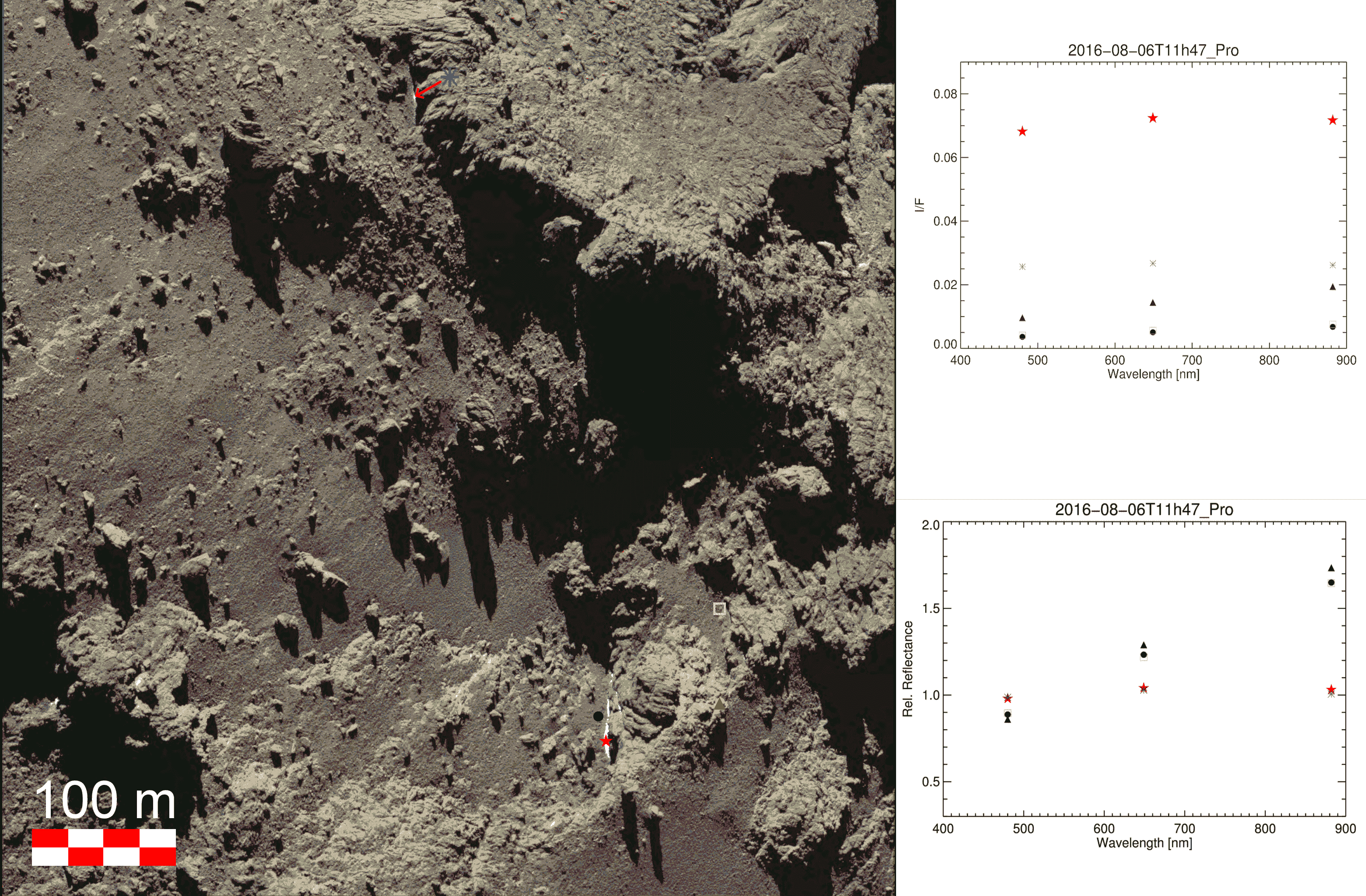}
 \end{center}
 \caption{RGB reflectance properties of BF-06 and of another neighboring bright
    surface as observed in the NAC 2016-08-06T11:47 sequence of observations. 
    The red star and blue asterisk indicate the bright feature at 
    the bottom of an underhang close to UR-00 and BF-06, respectively. Other symbols mark 
    neighboring smooth surfaces and consolidated material. In these observations, 
    the phase angle is 71.4\dg and the resolution is 16 cm/px. For clarity, 
    the area visible in this image is roughly delimited by the red square in fig. 
    \ref{fig: Flyby 2016 I/F}.}
 \label{fig: august 2016}
\end{figure*}

The OSIRIS/NAC observations taken during the April 2016 low-altitude low-phase 
angle flyby show in great detail the complexity of the transition between the 
morphological regions Imhotep and Khepry. While this area is similar to most of 
the overall nucleus surface, it contains a variety of morphological features as 
well as a diversity of behaviors in terms of colors and spectral properties.%
\\
Moreover, it shows some striking differences with the part of the transition 
area between the Ash and Imhotep regions that the spacecraft crossed during the 
February 2015 flyby \citep{Feller_2016} and that was surveyed with a 12 cm/px 
resolution.\newline\\
In the Ash-Imhotep transition area of the February 2015 flyby (see also table 
\ref{tbl: flyby comparison}), the part belonging to the Ash morphological unit 
appeared to be coated with fine-material deposits and peppered with decimeter- 
and meter-sized boulders, while the 400 m slope, leading toward the center of 
the Imhotep depression, consisted of strata of consolidated material covered 
with sparse boulders and a few small localized surfaces where fine material 
and pebbles are visible.\\
On the other hand, the Imhotep-Khepry transition presents a terraced topography 
where several extended fine-material deposits are visible on Imhotep, Khepry, 
and part of Aten. It also hosts large features of consolidated material, as 
well as areas of degraded outcrops, diamicton, taluses, gravitational 
accumulation deposits, and boulder fields.\newline\\
These two transitions further differ from one another in their global 
spectrophotometric properties. Some degree of difference was already indicated 
by August 2014 observations where at $\sim$ 1.3\dg of phase angle, the Ash-%
Imhotep transition exhibited reflectances at 649nm lower than 5.1\%, and a 
553-882 nm spectral slope higher than 13.5 \%/100 nm, whereas the Imhotep-%
Khepry transition presented reflectances of about 5\% and higher, and 
corresponding spectral slopes between 11 and 14 \%/100 nm (see fig. 9 in 
\citealp{Fornasier_2015}). Similarly, in fig. 13 of \citet{Fornasier_2015}, at 
around 10\dg\   phase angle, the former transition only displayed strongly 
elevated spectral slope values, while the latter presented a range of moderate 
and elevated spectral slopes values. Based on these observations, the Ash-%
Imhotep transition could then be classified as belonging to a group of 
terrains with high spectral slopes, and the Imhotep-Khepry to the group of 
terrains with intermediate spectral slopes.\newline\\
This group of terrains also included the regions close to the top of the small 
lobe of the comet, such as Agilkia. This particular region, investigated in 
\cite{LaForgia_2015}, shows clear variations in the assembled normal albedo and 
spectral slope mappings. As indicated by the authors, the comparison of these 
mappings with the corresponding morphological map points to a correspondence 
between the reflectance, spectral slope, and morphological nature of a terrain. 
In particular, it was observed that around the Agilkia area, the fine-material 
deposits exhibited on average a slightly higher reflectance and a lower 
spectral slope (computed following \cite{Jewitt_1986} in the 480-882nm range) 
than their locally surrounding diamicton fields, taluses, gravitational 
accumulation deposits, or neighboring outcropping layered terrains.\\
As apparent in the mappings of the previous section, a similar observation can 
be made for the Imhotep-Khepry transition. While in the reflectance mappings 
uneven and rough surfaces (e.g outcropping stratified terrains or degraded 
outcroppings) appear only slightly darker in this area than fine-material 
deposits, a clearer contrast arises from the spectral slope mappings in which 
smooth areas of fine-material deposits stand out from their rougher 
surroundings (e.g., diamictons or taluses). Other outstanding features add 
themselves to this general picture, such as the somber boulders, bright 
surfaces in niches, or close to underhangs and the large consolidated material 
feature encircled in red in the bottom panel of fig. 
\ref{fig: Flyby 2016 panels}, which presents lower-than-average spectral slopes 
at its head and flanks.\newline\\
While investigations into the nature of somber boulders are still ongoing, 
the bright surfaces displaying low spectral slopes that are essentially found 
in niches or at the bottom of underhangs, exhibit a behavior that is typically 
associated with areas that are enriched in water-ice material, as explained 
previously.\\
Any enrichment in water-ice material would be only fractional, however, as 
discussed in previous studies (\citealp{Sunshine_2006,Filacchione_2016a,
Barucci_2016,Oklay_2016} and \citealp{Fornasier_2016}). In particular, the 
analysis of surfaces observed by the VIRTIS infrared spectrometer, in which 
bright surfaces were visible, led by \citet{Raponi_2016}, has indicated that 
areas presenting a spectral slope lower than 10 \%/100 nm in the 500-1000 nm 
range at a phase angle of 95\dg\  could be composed of just over 1\% of pure 
water-ice in an areal mixing scenario and of well over 5\% in an intimate 
mixture scenario.\\
Additionaly, the area denoted by a blue circle in fig. \ref{fig: RGB GEOTIFF}, 
where bright surfaces have been identified by \cite{Pommerol_2015}, was 
considered to host material enriched by up to 6\% in water-ice according to 
infrared spectrum modeling \citep{Filacchione_2016a}. For some of the surface 
elements of the same area, \citet{Oklay_2017} have obtained corresponding 
values ranging from 6\% to 25\% based on the spectrophotometric properties of 
these spectral regions and thermal modeling. Similarly, the analyses conducted 
in the Anhur and Bes on some of the singular very large compositional 
heterogeneities have indicated local $\sim$ 20\% and $\sim$ 30\% enrichment in 
water ice (\citealp{Fornasier_2016} and \citealp{Fornasier_2017}).\newline\\
At this time, we cannot report any constraint on the water-ice enrichment of 
the observed bright surfaces from the considered dataset. We can report, 
however, that some of the bright surfaces have been observed several times 
during the mission and that investigations of a temporal variation in their 
spectrophotometric properties is currently being undertaken. For instance, we 
have found the bright feature BF-06 and another bright feature close to UR-0 to 
clearly present once again in August 2016 at 3.5 au a spectral behavior that is 
consistent with a water-ice material enrichment (see fig. 
\ref{fig: august 2016}).
In this observation, the reflectance of the bright surfaces is clearly between 
4 and 10 times higher than that of neighboring smooth surfaces of fine-material 
deposits, and it clearly exhibits a spectral behavior with a neutral slope with 
respect to the same neighboring surfaces. This observation therefore indicates 
the persistence of these bright surfaces, and in the paradigm of a spectral 
behavior consistent with a water-ice material enrichment, the lack of 
sufficient insolation to sublimate this material between April and August 2016. 
As the comet was at 3.5 au, it would then be quite likely that these surfaces 
would survive until the next approach to perihelion.\newline\\
In the bright surfaces investigated conjointly by the OSIRIS and VIRTIS 
instruments, water-ice material is absolutely certain. However, the exact 
nature of the average terrain material that surrounds these bright features 
and composes the top-layer of the nucleus surface (also referred to as “dark 
terrain” in \citealp{Fornasier_2017}) still remains to be defined.\\
Comparison of the photometric results from the February 2015 flyby with 
laboratory measurements has found that intra-mixture of carbon, tholins, and 
water ice \citep{Jost_2017a} matched the phase curve better than the inter-%
mixture counterpart. A similar observation was made regarding the nature of 
bright spots for this flyby observed with the OSIRIS/WAC camera 
\citep{Hasselmann_2017}.\\ 
However, the comparison of these samples with the measurements of smooth fine-%
material deposits of the Ash-Imhotep transition had indicated a mismatch of 
these particular samples in terms of the spectral slopes, as tholinsinduced a 
stronger spectral slope than had been observed (see fig. 16 of 
\citealp{Feller_2016}).\newline\\
Recent researches into the spectrophotometric properties of cometary analogs 
have been notably reported in \citet{Jost_2017a, Jost_2017} and 
\citet{Rousseau_2017}. While the first study furthered the analysis of carbon, 
tholins, and water-ice mixtures, the latter investigated the properties of 
coal mixed with silicates and/or with pyrrhotites (iron sulfides).\\
In both studies, the authors have found mixtures that provided a satisfactory 
match of the comet spectrum in the 250-1000nm range. Additionally, 
\citet{Jost_2017} also reported mixtures that matched the observed phase-%
reddening phenomenon. However, \citet{Jost_2017} also noted that the 
investigated mixtures did not provide a simultaneous match of the comet albedo, 
spectrum, phase curve, and of the phase reddening.\\
Furthermore, \citet{Rousseau_2017} noted that although they found mixtures that
matched the comet reflectance, these mixtures did not match the observed 
spectral slope of the average comet terrain either. They pointed out that the 
organic compound they used induced a lower spectral slope than the one of the 
comet.\newline\\
Consequently, investigations for an appropriate cometary analog for the surface 
of 67P in terms of spectrophotometric and physical properties, as pointed in 
part in \cite{Kaufmann_2018}, for instance, are still ongoing.\\
In addition to considering the albedo and reflectance spectrum of a mixture, 
including phase-reddening measurements might prove a pertinent additional 
comparison factor. However, the grounds for this phenomenon are still the 
subject of research and remain to be well defined. This phenomenon has 
tentatively been attributed to the increased contribution of the multiple 
scattering at large phase angles as the wavelength and albedo increase, and 
to a contribution of surface roughness effects (\citealp{Hapke_2012, 
Sanchez_2012} and \citealp{Schroeder_2014}). \citet{Jost_2017a} further 
underlined that the nature of a surface phase reddening also appears to depend 
on its composition, and that in their surface analog mixtures, both the 
darkening agent (carbon particles) and the organic compound (tholins) had a 
distinctive influence on the presence and strength of the phase reddening.\\
While additional theoretical investigation and laboratory experiment 
validation are crucial to determine the frame of the phase-reddening 
phenomenon within a low temperature, irradiated and Van-der-Waals forces 
dominated environment, considering the phase reddening in current laboratory 
experiments would nevertheless constitute a further valuable comparison tool 
for determining an appropriate analog of the surface of 67P.
\newline\\
\section{Conclusions}
We have presented here the results of the geomorphological mapping and of the 
spectrophotometric analysis based on the OSIRIS/NAC images acquired during the 
April 2016 flyby manoeuvre above the Imhotep-Khepry transition area on 67P.
\begin{itemize}
 \item We have identified and mapped the host of morphological units that are 
   visible in the Khepry-Imhotep transition, which, without marking this region 
   as unique, indicate its diversity and relative complexity.
 \item We have performed spectrophotometric analyses of the transition in 
   general as well as of peculiar surface elements. While the smooth-looking 
   regolith of this region is similar to the average dark terrain of the nucleus
   in terms of reflectance and spectral slope, rough terrains such as 
   diamictons, degraded outcrops, and consolidated material exhibit a lower-%
   than-average reflectance and a higher-than-average spectral slope.\\
   One particular outcrop of consolidated material roofed by a cuesta presents 
   a slightly higher-than-average reflectance and a slightly lower-than-average 
   spectral slope, which likely indicates a local compositional heterogeneity 
   at a scale of some tens of meters.\\
   Additionally, some meter-sized features also present a peculiar 
   spectrophotometric behavior. Sombre boulders, which are ubiquitous in the 
   Imhotep-Khepry transition, show a similar spectral behavior as the 
   neighboring smooth terrains, and their reflectance is $\sim$20\% lower than 
   average. The bright surfaces of this area are unmistakable, their 
   reflectance is at least at twice the local average, and their spectral 
   behavior is repeatedly less steep than the average surface. These features 
   both indicate small-scale compositional heterogeneities across the surface 
   of this region.
 \item We have identified that the bright features that exhibit moderate to 
   neutral spectral slopes are likely candidates for locations that are 
   enriched in water-ice material. We have also shown that some of the bright 
   surfaces investigated in April 2016 that are likely enriched in water-ice 
   material have persisted up to and beyond the frost line, and at least up 
   to August 2016. This further supports evidence for long-term survival of 
   water-ice-rich material and frosts on the nucleus surface.
 \item We have measured the local phase-reddening parameters as the comet was 
   at 2.7 au outbound from perihelion, and have found them to be on the same 
   order as those obtained for the Ash-Imhotep transition observed during the 
   February 2015 flyby, which further indicates seasonal variations on the 
   surface of the nucleus.
\end{itemize}
Accordingly, this transition area displays a diversity in reflectances and 
spectral slopes that is as great as it is in morphological terms. This 
indicates that the variety in the surface top-layer composition was greater 
than in the flyby area on 14 February 2015.

% Acknowledgments section
\section*{Acknowledgements}
OSIRIS was built by a consortium of  the Max-Planck-Institut f\"ur
Sonnensystemforschung, G\"ottingen, Germany, CISAS--University of Padova,
Italy, the Laboratoire d'Astrophysique de Marseille, France, the Instituto de
Astrof\'isica de Andalucia, CSIC, Granada, Spain, the Research and Scientific
Support Department of the European Space Agency, Noordwijk, The Netherlands,
the Instituto Nacional de T\'ecnica Aeroespacial, Madrid, Spain, the Universidad
Polit\'echnica de Madrid, Spain, the Department of Physics and Astronomy of
Uppsala University, Sweden, and the Institut  f\"ur Datentechnik und
Kommunikationsnetze der Technischen Universit\"at  Braunschweig, Germany. The
support of the national funding agencies of Germany (DLR), France (CNES), Italy
(ASI), Spain (MEC), Sweden (SNSB), and the ESA Technical Directorate is gratefully
acknowledged.\\
Rosetta is an ESA mission with contributions from its member states and NASA.
Rosetta's Philae lander is provided by a consortium led by DLR, MPS, CNES and
ASI.\\
The SPICE libraries and PDS resources are developed and maintained by NASA.\\
The authors thanks the referee and editors for their questions, remarks and 
advices for the improvement of this manuscript.

%%%%%%%%%%%%%%%%%%%%%%%%%%%%%%%%%%%%%%%%%%%%%%%%%%

\bibliographystyle{astron_tmp}
\bibliography{BiblioJabref}
%%%%%%%%%%%%%%%%%%%%%%%%%%%%%%%%%%%%%%%%%%%%%%%%%%

%%%%%%%%%%%%%%%%% APPENDICES %%%%%%%%%%%%%%%%%%%%%

\appendix
\section{Additional material}
\begin{table*}
 \centering
  \begin{tabular}{|c|c|c|c|c|c|c|c|}
   \hline
   \hline
   April 2016 (UTC) & NAC Filters & D (km) & Res. (m/px)& $\alpha_{m}$ (\dg) & $\alpha_{min}$ - $\alpha_{max}$  (\dg) & Longitude  & Latitude\\
   \hline
   09T12:15:17.916 & All filters   & 47.0$\pm$0.6 & 0.87$\pm$0.05 & 60.6 & 59.5-61.7 & 149.89\dg & -5.76\dg \\
   09T23:01:09.919 & F84, F82, F88 & 28.8$\pm$0.2 & 0.54$\pm$0.02 & 7.80 & 6.67-8.94 & 121.33\dg &-17.09\dg \\
   09T23:08:09.910 &  ”            & 28.8$\pm$0.2 & 0.54$\pm$0.02 & 6.99 & 5.85-8.14 & 118.12\dg &-16.77\dg \\
   09T23:14:09.894 &  ”            & 28.8$\pm$0.2 & 0.54$\pm$0.02 & 6.29 & 5.16-7.45 & 115.04\dg &-16.64\dg \\
   09T23:19:08.951 &  ”            & 28.8$\pm$0.2 & 0.53$\pm$0.02 & 5.72 & 4.58-6.87 & 112.54\dg &-16.50\dg \\
   09T23:22:09.905 &  ”            & 28.7$\pm$0.1 & 0.53$\pm$0.02 & 5.37 & 4.23-6.53 & 111.24\dg &-16.29\dg \\
   09T23:25:09.886 &  ”            & 28.7$\pm$0.1 & 0.53$\pm$0.02 & 5.02 & 3.88-6.19 & 109.92\dg &-16.11\dg \\
   09T23:28:09.912 &  ”            & 28.7$\pm$0.1 & 0.53$\pm$0.02 & 4.68 & 3.53-5.85 & 108.70\dg &-15.87\dg \\
   09T23:31:09.942 &  ”            & 28.7$\pm$0.1 & 0.53$\pm$0.01 & 4.34 & 3.19-5.50 & 107.51\dg &-15.62\dg \\
   09T23:34:09.904 &  ”            & 28.7$\pm$0.1 & 0.53$\pm$0.01 & 3.99 & 2.84-5.18 & 106.47\dg &-14.71\dg \\
   09T23:37:09.883 &  ”            & 28.7$\pm$0.1 & 0.53$\pm$0.01 & 3.65 & 2.48-4.88 & 105.35\dg &-13.30\dg \\
   09T23:40:09.927 &  ”            & 28.6$\pm$0.1 & 0.53$\pm$0.01 & 3.30 & 2.10-4.58 & 104.30\dg &-11.19\dg \\
   09T23:43:09.884 &  ”            & 28.6$\pm$0.1 & 0.53$\pm$0.01 & 2.94 & 1.69-4.26 & 103.33\dg & -8.71\dg \\
   09T23:46:09.901 &  ”            & 28.6$\pm$0.1 & 0.53$\pm$0.01 & 2.57 & 1.25-3.93 & 102.74\dg & -5.78\dg \\
   09T23:49:09.982 &  ”            & 28.6$\pm$0.1 & 0.53$\pm$0.01 & 2.19 & 0.80-3.56 & 103.48\dg & -2.47\dg \\
   09T23:52:10.005 &  ”            & 28.6$\pm$0.1 & 0.53$\pm$0.01 & 1.82 & 0.39-3.17 & 104.65\dg &  0.87\dg \\
   09T23:55:09.984 &  ”            & 28.6$\pm$0.1 & 0.53$\pm$0.01 & 1.48 & 0.095*-2.79 & 106.05\dg &  3.32\dg \\
   09T23:58:09.911 &  ”            & 28.6$\pm$0.1 & 0.53$\pm$0.01 & 1.24 & 0.095*-2.47 & 107.85\dg &  5.11\dg \\
   10T00:01:09.929 &  ”            & 28.6$\pm$0.1 & 0.53$\pm$0.01 & 1.21 & 0.095*-2.43 & 110.21\dg &  5.88\dg \\
   10T00:04:09.927 &  ”            & 28.6$\pm$0.1 & 0.53$\pm$0.01 & 1.42 & 0.095*-2.72 & 112.57\dg &  5.78\dg \\
   10T00:07:10.009 &  ”            & 28.6$\pm$0.1 & 0.53$\pm$0.01 & 1.75 & 0.31-3.09 & 114.98\dg &  4.60\dg \\
   10T00:10:09.944 &  ”            & 28.7$\pm$0.1 & 0.53$\pm$0.01 & 2.11 & 0.70-3.49 & 117.11\dg &  2.31\dg \\
   10T00:13:10.092 &  ”            & 28.7$\pm$0.1 & 0.53$\pm$0.01 & 2.48 & 1.14-3.85 & 118.13\dg & -0.52\dg \\
   10T00:16:08.998 &  ”            & 28.8$\pm$0.1 & 0.53$\pm$0.01 & 2.84 & 1.57-4.18 & 118.36\dg & -3.40\dg \\
   10T00:19:09.893 &  ”            & 28.8$\pm$0.1 & 0.54$\pm$0.01 & 3.19 & 1.98-4.48 & 117.89\dg & -6.17\dg \\
   10T00:22:09.912 &  ”            & 28.8$\pm$0.2 & 0.54$\pm$0.01 & 3.54 & 2.36-4.78 & 116.89\dg & -8.22\dg \\
   10T00:25:09.908 &  ”            & 28.9$\pm$0.2 & 0.54$\pm$0.02 & 3.88 & 2.72-5.08 & 115.50\dg & -9.78\dg \\
   10T00:28:09.931 &  ”            & 28.9$\pm$0.2 & 0.54$\pm$0.02 & 4.21 & 3.06-5.38 & 113.83\dg &-10.48\dg \\
   10T00:31:09.977 &  ”            & 28.9$\pm$0.2 & 0.54$\pm$0.02 & 4.54 & 3.40-5.71 & 112.01\dg &-10.51\dg \\
   10T00:34:09.905 &  ”            & 28.9$\pm$0.2 & 0.54$\pm$0.02 & 4.88 & 3.73-6.04 & 110.17\dg &-10.44\dg \\
   10T00:37:09.911 &  ”            & 28.9$\pm$0.2 & 0.54$\pm$0.02 & 5.21 & 4.07-6.37 & 108.31\dg &-10.26\dg \\
   10T00:40:09.924 &  ”            & 29.0$\pm$0.2 & 0.54$\pm$0.02 & 5.54 & 4.40-6.70 & 106.39\dg & -9.98\dg \\
   10T00:43:09.896 &  ”            & 29.0$\pm$0.2 & 0.54$\pm$0.02 & 5.87 & 4.74-7.03 & 104.49\dg & -9.66\dg \\
   10T00:46:09.944 &  ”            & 29.0$\pm$0.2 & 0.54$\pm$0.02 & 6.21 & 5.07-7.46 & 102.60\dg & -9.41\dg \\
   10T11:50:16.568 & All filters   & 53.8$\pm$0.5 & 1.00$\pm$0.05 & 51.2 & 49.9-52.2 &  87.31\dg & -14.82\dg \\
   \hline
  \end{tabular}
 \caption{\label{tab: Flyby 2016 NAC images} List of OSIRIS/NAC observations 
    used in this study: in the table above, D is the median distance between 
    spacecraft to imaged surface, $\alpha_{m}$ is the median phase angle, and 
    the next column gives the amplitude of the phase angle values. The 
    longitudes and latitudes are those of the NAC boresight in the Cheops frame 
    (see the main text for details).
    }
\end{table*}
\begin{table*}
 \centering
 \begin{tabular}{|c|c|c||c|c|c||c|c|c|}
  \hline
  \hline
  Feature & Long.    & Lat.     & 23h34        &  23h46       & 00h01        & 23h34          &  23h46      & 00h01       \\
 \hline
  $ $     & $\pm$0.1 & $\pm$0.1 & \multicolumn{3}{|c|}{I/F at 649 nm (\%)}     & \multicolumn{3}{|c|}{Spectral slope (\%/100 nm)}\\
  \hline
  UR-00  &  -46.7   &   -8.3   &  5.3$\pm$0.1 &  5.6$\pm$0.1 &  6.4$\pm$0.1 & 17.7$\pm$0.2  & 18.0$\pm$0.5 & 18.3$\pm$0.2\\
  BF-01  &  -56.5   &   -8.3   & 20.0$\pm$3.0 &     $   $    &     $   $    &  8.0$\pm$11.0 &     $   $    &     $   $   \\
  BF-02  &  -43.1   &  -27.8   &  9.0$\pm$0.6 &     $   $    &     $   $    &  7.0$\pm$1.5  &     $   $    &     $   $   \\
  BF-03  &  -46.0   &  -20.3   &  7.0$\pm$0.2 &  7.6$\pm$0.1 &     $   $    & 11.0$\pm$2.0  & 10.8$\pm$0.4 &     $   $   \\
  BF-04  &  -45.9   &  -20.4   &     $   $    &  8.8$\pm$0.3 &     $   $    &     $   $     &  8.5$\pm$0.6 &     $   $   \\
  BF-05  &  -45.7   &  -12.6   &  9.0$\pm$1.6 & 11.0$\pm$1.4 & 11.0$\pm$1.2 &  5.1$\pm$5.0  &  6.0$\pm$2.0 &  9.0$\pm$3.0\\
  BF-06  &  -41.8   &   -6.9   &  9.0$\pm$1.0 & 11.0$\pm$1.2 & 14.0$\pm$1.7 &  7.0$\pm$3.0  &  5.0$\pm$5.0 &  5.0$\pm$4.0\\
  BF-07  &  -39.0   &   -6.7   &  6.4$\pm$0.2 &  7.0$\pm$0.3 &  8.0$\pm$0.3 & 15.9$\pm$0.5  & 16.2$\pm$0.6 & 15.0$\pm$1.0\\
  BF-08  &  -51.6   &   -4.3   & 12.0$\pm$1.3 & 11.0$\pm$1.1 &     $   $    &  2.3$\pm$1.5  &  4.0$\pm$1.1 &     $   $   \\
  BF-09  &  -54.2   &   -3.3   &     $   $    &     $   $    &  8.1$\pm$0.1 &     $   $     &     $   $    & 10.6$\pm$0.5\\
  BF-10  &  -32.5   &   13.2   &     $   $    &     $   $    & 11.3$\pm$0.6 &     $   $     &     $   $    &  6.8$\pm$0.6\\
  SB-00  &  -51.4   &  -24.9   &  4.2$\pm$0.1 &     $   $    &     $   $    & 20.1$\pm$0.7  &     $   $    &     $   $   \\
  SB-01  &  -43.9   &  -18.4   &  4.9$\pm$0.1 &  5.3$\pm$0.1 &     $   $    & 20.2$\pm$0.5  & 20.1$\pm$0.7 &     $   $   \\
  SB-02  &  -45.5   &  -11.7   &  4.8$\pm$0.1 &  5.2$\pm$0.1 &     $   $    & 21.1$\pm$0.5  & 21.8$\pm$0.4 &     $   $   \\
  SB-03  &  -27.7   &   -3.8   &  5.3$\pm$0.1 &  5.8$\pm$0.1 &     $   $    & 18.4$\pm$0.3  & 18.5$\pm$0.1 &     $   $   \\
  SB-04  &  -20.8   &    7.0   &  5.0$\pm$0.1 &  5.4$\pm$0.1 &     $   $    & 18.5$\pm$0.2  & 18.8$\pm$0.1 &     $   $   \\
  SB-05  &  -20.0   &    7.5   &  5.1$\pm$0.1 &  5.5$\pm$0.1 &  5.7$\pm$0.1 & 19.5$\pm$0.3  & 19.7$\pm$0.2 & 19.5$\pm$0.2\\
  SB-06  &  -42.6   &    6.5   &     $   $    &  5.0$\pm$0.1 &  5.6$\pm$0.1 &     $   $     & 19.7$\pm$0.5 & 18.5$\pm$0.4\\
  SB-07  &  -17.4   &    9.6   &     $   $    &     $   $    &  5.4$\pm$0.1 &     $   $     &     $   $    & 19.0$\pm$0.1\\
  \hline
 \end{tabular}
 \caption{\label{tbl: spectral slopes tbl} Extended list of features investigated:
    In the table above, the first column indicates the ID of the 
    feature, the second and third columns give the features' positions 
    relative to the Cheops boulder, the two sets of three columns list their 
    reflectance at 649 nm and their spectral slopes in the 535-743 nm range 
    for the three sequences of observations acquired 
    close to the moment of opposition and discussed in the main text.
   }
\end{table*}
\begin{table*}
 \centering
 \begin{tabular}{|c|c|c|}
  \hline
  \hline
  Properties        & February 2015   & April 2016      \\
  \hline                                                
  \# NAC images     & 91              & 121             \\
  Filters used      & 24 sequences with F84,F82,F88  & 33 sequences with F84,F82,F88  \\
  $ $               & 2 sequences with all filters   & 2 sequences with all filters   \\ 

  Sequences cadence (seconds)         & 36 - 600        & 180 - 420       \\
  Best pixel scale  & 0.11 m/pxl      & 0.53 m/pxl      \\
  Phase angle range & 0\dg - 33\dg    & 0\dg-12\dg + $\sim$ 51\dg + $\sim$ 60\dg \\
  \hline
  Phase reddening   & S(0\dg) = 17.9\%/100 nm             & S(0\dg) = 17.9\%/100 nm             \\
  parameters        & $\beta$ = 6.52 $\cdot$ 10$^-4$/100nm/\dg & $\beta$ = 6.40 $\cdot$ 10$^-4$/100nm/\dg \\
  \hline
  Median p$_{v}$(0\dg, 649 nm)        & 6.3\%$\pm$0.6\% & 6.3\%$\pm$0.6\%   \\
  Max p$_{v}$(0\dg, 649 nm)           & 10.4\%           & 16.0\%* \\
  \hline
  Peculiar features & Strata heads,   & bright patches, \\
     $ $            & bright patches, & somber boulders \\
     $ $            & somber boulders &  $   $          \\
  \hline
  Investigated surfaces & $  $        &  $   $   \\
  Bright patches        & Spectral slope$^{\dagger}$ > 17\%/100nm & Spectral slope$^{\dagger}$ < 10\%/100nm \\
  Sombre boulders       & Spectral slope$^{\dagger}$> 17\%/100nm & Spectral slope$^{\dagger}$ > 17\%/100nm    \\
  \hline
  Spectral differences  & Slight      & Strong   \\
  (bright/somber features) & $  $     & $  $     \\
  \hline
  Bright features' spectral     & average dark terrain    & water-ice rich material     \\
  behavior similar to            &    $  $                 &     $  $                    \\
  \hline
  Favored locations of bright features & None obvious  & Bottom of niches and underhangs \\ 
  \hline
 \end{tabular}
 \caption{\label{tbl: flyby comparison} Comparison of the main characteristics 
    of the February 2015 and April 2016 flybys: Abridged comparative of the two 
    flybys discussed in     \citet{Feller_2016} and in the main text.\\
    $\dagger$: Spectral slope close to 0\dg of phase angle.
    }
\end{table*}

% If you want to present additional material which would interrupt the flow of 
% the main paper, it can be placed in an Appendix which appears after the list 
% of references.
%%%%%%%%%%%%%%%%%%%%%%%%%%%%%%%%%%%%%%%%%%%%%%%%%%

\end{document}